\documentclass[useAMS,usenatbib]{mn2e}
\usepackage[pdftex,pdfpagemode={UseOutlines},bookmarks,bookmarksopen,
colorlinks,linkcolor={blue},citecolor={blue},urlcolor={red}]{hyperref}

 \usepackage{graphicx}  
\usepackage{amsmath}
\usepackage{soul}
\usepackage{color}
\usepackage{ifpdf}
\usepackage{epsfig}

\def\aj{AJ}
\def\apj{ApJ}
\def\apjl{ApJ}
\def\apjs{ApJS}
\def\aap{A\&A}
\def\aaps{A\&AS}
\def\aapr{A\&AR}
\def\mnras{MNRAS}
\def\nat{Nature}
\title[Stellar populations of dE nuclei]{Nuclei of
  early-type dwarf galaxies: insights from stellar
  populations\thanks{Based on observations collected at the European
    Organisation for Astronomical Research in the Southern Hemisphere,
    Chile (programme 078.B-0178)}}

\author[Paudel et al.]{Sanjaya Paudel$^{1}$\thanks{E-mail:
    sjy@x-astro.net}, Thorsten Lisker$^{1}$, Harald
  Kuntschner$^{2}$\\
 $^{1}$Astronomisches Rechen-Institut, Zentrum f\"ur Astronomie der
  Universit\"at Heidelberg, M\"onchhofstr.\ 12-14, 69120
  Heidelberg, Germany\\
 $^{2}$Space Telescope European Coordinating Facility, European
 Southern Observatory, Karl-Schwarzschild-Str. 2, 85748 Garching, Germany}

\begin{document}

\date{  Accepted ...  Received ... ; in original form \today }

\pagerange{\pageref{firstpage}--\pageref{lastpage}} \pubyear{2010}

\maketitle

\label{firstpage}

\begin{abstract}
We present a comprehensive analysis of the spatially resolved stellar population
properties of 26 early-type dwarf (dE) galaxies in the Virgo
cluster. Using 
Lick/IDS absorption line indices we derive
simple stellar population(SSP)-equivalent age, metallicity
and [$\alpha$/Fe] abundance ratio. In particular, we focus on the
comparison of the stellar populations 
between the central nucleus and the surrounding galactic main
body. The stellar populations of the nuclei are, for most dEs, significantly
younger than those of the respective galactic main bodies, with an
average difference of 3.5 Gyr. We find only five dEs with
significantly older nuclei than their galactic main
bodies. Furthermore, we observe most dE nuclei to be more metal rich
compared to their host galaxies. These age and metallicity behaviours
are shown by almost all dEs brighter
than M$_{\it r}$ = -17 mag.

The
metallicity of both nuclei and galactic main bodies correlates
with the total luminosity of the dEs. However, the metallicity of the
nuclei covers
a larger range (+0.18 to -1.22 dex) than that of the galactic main
bodies, which all have sub-solar metallicity.
The ages of dE nuclei show a statistically significant correlation
with the local projected galaxy density within the cluster, such that
younger ages are predominantly observed outside of the high-density
central cluster region.
The
alpha-element abundance ratios are consistent with solar for both
nuclei and galactic main bodies.

 We
also examine the presence of radial gradients in the SSP parameters for a subset of 13 dEs
(up to 1.2 kpc or 15
arcsec radius). We notice two different types
of gradients, namely smooth profiles that include the nucleus, and
profiles where a break occurs between the nucleus and the rest of the
galaxy. Nevertheless, an overall trend of increasing age and
decreasing metallicity with radius exists, consistent with earlier
studies. The $\alpha$-abundance ratio as function of radius is
consistent with no gradient.

Possible formation scenarios for the nuclei of dEs are discussed. The
young and metal-enhanced population of nuclei suggests that these
might have formed at later epochs, or the termination of star formation
activity in the nuclei might have occured relatively late, perhaps due
to continuous infall
of gas into the central potential well. Our stellar population
analysis suggests that the merging of globular clusters is not an
appropriate scenario for the formation of most dE nuclei, at least not
for the brighter dEs. We speculate that there might be
different formation processes which are responsible for the formation
of dEs and their nuclei depending on their luminosity.

\end{abstract}


\section{Introduction}

Early-type dwarf galaxies (dEs, M$_{\it B}$ $>$ -18) are the
numerically dominant population in the present-day Universe
\citep{Sandage85, Binggeli87, Ferguson94}. They also exhibit  
strong clustering, being found predominantly in the close vicinity of
giant galaxies, either as satellites of individual giants, or as
members of galaxy clusters \citep{Ferguson89}. Although the dEs are
characterized by their smooth appearance, having no recent or ongoing
star formation and apparently no gas or dust content, the
understanding of their origin and evolution remain major challenges
for extragalactic astronomy. Stellar 
population studies show that dEs exhibit on average younger
ages as compared to their giant counterparts, and also a lower metal
content according to the correlation of metallicity and luminosity
\citep{Michielsen08}. However, past studies
 provided a wide range of ages (e.g.,
\citealt{Poggianti01}; \citealt{Rakos01}; \citealt{Caldwell03};
\citealt{Geha03}; \citealt{van04}), from as old as being primordial objects to 
dEs with recently formed young stellar populations.

It appears that dEs themselves are not a homogeneous class of objects. Sub-structures such as stellar disks, 
faint spiral arms or
bars are quite frequent among the brighter dEs
\citep{Lisker06a,Lisker07}. Many dEs were found to contain a central
surface brightness enhancement consistent with a point source on top
of the galactic main body \citep[e.g.][]{Binggeli91, Binggeli93}, 
referred to as so-called nucleated dEs.
The studies from the HST/ACS Virgo cluster survey \citep{Cote06}, with
their high angular resolution, not only
verified the presence of such a distinct nucleus but also showed that nuclei are
ubiquitous in bright dEs, covering a range in nucleus
brightness. Interestingly, dEs with comparably faint nuclei that had not been
identified before \citet{Cote06} show several systematically different
properties as compared to dEs with bright nuclei \citep{Lisker07,Lisker08}.

Different studies of dE nuclei from different data sets
found several contradictory properties for the nuclei
\citep{Grant05,Lotz04b,Cote06}. Particularly, the ground and space
based data sets yielded different results. \citet{Grant05} found
that the nuclei are on average redder than their surrounding galactic main
body. On the other hand, studies using HST
observations \citep{Cote06,Lotz04b}
measured the dE nuclei to be slightly bluer than the galactic
part. Furthermore,  \citet{Cote06}, who used high quality data sets
from the ACS Virgo Cluster Survey, proposed that the nuclei rather
closely match the nuclear clusters of late type spiral galaxies in terms of size,
luminosity and overall frequency. Another related scenario
is also emerging: the recently discovered new (candidate) type of extremely
small dwarf galaxies, the UCDs (Ultra Compact Dwarfs) with typical
magnitudes of  $-13 < M_b < -11$ \citep{Hilker99,Phillipps01}, might
be the remnant nuclei of tidally stripped dwarf galaxies
\citep{Bekki03,Drinkwater03,Goerdt08}. 

The formation mechanisms of the nuclei of dEs are poorly
understood and various possibilities have been proposed, also
depending on the  evolution and formation of dEs as a whole. As the
nucleated dEs are preferentially rounder in shape, \citet{Bergh86}
proposed that the nuclei of dEs could have formed from the gas that
sank to the centre of the more slowly rotating objects. Since they
predominantly appear in highly dense environments, like the centre of
a cluster of galaxies, the pressure from the surrounding
inter-galactic medium may allow dwarf galaxies to retain their gas
during star formation and produce multiple generation of stars
\citep{Silk87,Babul92}, forming nuclei in the
process. In both proposed scenarios the nuclei are formed along with the
evolution of the galaxy itself, i.e., continuous star formation
activity occurs at the dE centre as time passes. Unlike that, \citet{Oh00}
suggested that dE nuclei might have formed in a different way, namely
through subsequent migration or orbital decay of several globular
clusters towards the centre of their host dE.

It is difficult to provide a definitive observational test of
these different scenarios for nucleus formation. Nevertheless, we
can gain some insight by comparing the different observational
properties, in particular relative ages and chemical enrichment
characteristics, of the nuclei with their galactic main bodies, as
well as with UCDs as their possible descendants. However, we need to
bear mind that there may be a mixture of different formation scenarios.

Our previous study based on this dataset \citep[hereafter Paper I]{Paudel10} has focused on the analysis of
the inner
stellar populations of dEs as a whole, without separating nuclei and
galactic main bodies. Instead, our intention was to see the variation
of the inner
stellar population properties with different morphological subclasses
of dEs \citep[cf.][]{Lisker07}, using a much larger sample of
Virgo dEs than in previous Lick index studies. We showed that dEs with
different substructure properties
(with/without disk features, \citealt{Lisker06a}) have significantly different stellar
populations:  dEs with disk features are younger and more metal
rich than dEs without disks. Therefore we concluded that these dEs
probably do not have the same origin, as they also differ in their
distribution with local environmental density in which they reside. By
selection, all dEs in our sample contain a central nucleus, therefore it seems important
to see the nature of the stellar populations of the nuclei and the
surrounding galactic main bodies separately. And since there are different
possibilities for the processes that form nuclei and also dEs
themselves, we ask: can the
nuclei thus tell us something about the formation history of dEs? 

This paper is organized as follows. In Section 2, we describe the sample of Virgo cluster dEs, observation and data 
reduction in brief. In section 3, we describe the measurement of line-strength indices in the Lick/IDS system. Our 
main results from the stellar population parameters are given in
Section 4 and are discussed in 
Section 5. Finally, we summarize our findings in Section 6.

\section{The Sample, Observation and Data Reduction} \label{sample}
Our sample comprises 26 nucleated dEs in the Virgo cluster. The
sample properties such as position in the color magnitude relation, total
galactic luminosity, radial velocity and their local projected density
within the Virgo cluster are described in detail in Paper I. The sample
covers the full range of local density and includes the different
morphological dE subtypes, i.e., 8 dEs with disks (dE(di)s) and 18 dEs
without disks, which we hereafter simply refer to as dE(N)s.  One 
dE(di) (VCC0308) contains a weak blue color excess in the centre,
thus being referred to as a blue-centre dE \citep[cf.][]{Lisker06b}.

The observations were carried out at the ESO Very Large Telescope (VLT)
with the FORS2 instrument. The 1" slit and 300V grism provide an
instrumental resolution of $\simeq$11 \AA{}(Full Width at Half Maximum, FWHM). The other basic
observational properties and the data reduction processes are
described in detail in Paper I.

We carefully checked the issue of
scattered light during the reduction of the data, since the presence
of a significant amount of scattered light could produce an artificial
gradient in the measured line indices. Fortunately, our MOS-MXU setup
utilized in this investigation provides the opportunity to quantify
it. There are always free intra-slit regions where no light enters
directly from the sky. After the bias subtraction these regions should
not contain any flux, unless scattered light were present.  We thus
calculate the average amount of light within such regions manually. We
find that the mean is zero within the uncertainties, which are of the
order of some hundredths of a count. The FORS2 pipeline reduction
produces the same result. It therefore confirms that there is no scattered
light left in the spectra.

In a different way, there is still the probability of mixing the nucleus
light out to far beyond the central nucleus in case of bad seeing or
instrumental blurring. To examine this effect, we also observed a
star in an additional slit along with each target-field. Then, through the light
profile of this star, we quantify the amount of such light 
at a radius of 3" beyond the centre. Our measurements show that spread nuclear light is less than 1\% of galactic 
light at 3" distance from the galaxy centre. The observed FWHM of the stars is always $\sim$1.3" or less,
consistent with this negligible fraction of starlight at 3" from the
centre.

\subsection{Extraction of nuclear spectra and analysis of light profile } \label{extrac}
\begin{table}
 \centering
\begin{minipage}{8cm}
  \caption{Basic parameters and signal-to-noise ratio for our targets.}
  \label{snr}
\begin{tabular}{lllllll}
\hline
Galaxy	&	Nuc. &	Gal.	&	Reff	&	M$_{\it r}$	& m$_{\it r}$	&Light \\
 {\tiny VCC}		&	{\tiny(SNR)}		&
 {\tiny(SNR)}	& arcsec	& {\tiny Total}	& {\tiny Nuc}  & fraction	\\
 No.		&	pix$^{-1}$	&	pix$^{-1}$	& arc-sec		&  mag &  mag & in  \%	\\		
\hline
0216 	&	47	&	30	&	13.3	&	$-$16.78	  &	  -11.58  &	22	\\
0308 	&	35	&	31	&	18.7	&	$-$17.95	  &	  -11.91  &	40	\\
0389 	&	32	&	30	&	17.2	&	$-$18.00	  &	  -12.59  &	48	\\
0490 	&	32	&	23	&	27.6	&	$-$18.09	  &	  -12.43  &	25	\\
0545 	&	33	&	30	&	13.3	&	$-$16.61	  &	  -11.71  &	35	\\
0725 	&	23	&	--	&	25.2	&	$-$16.19	  &	  -10.17  &	$-$	\\
0856 	&	56	&	35	&	15.9	&	$-$17.71	  &	  -12.73  &	23	\\
0929 	&	56	&	34	&	20.5	&	$-$18.58	  &	  -13.13  &	33	\\
0990 	&	35	&	33	&	09.9	&	$-$17.39	  &	  -12.52  &	53	\\
1167 	&	46	&	30	&	27.3	&	$-$16.95	  &	  -12.01  &	17	\\
1185 	&	33	&	50	&	19.3	&	$-$16.65	  &	  -10.76  &	30	\\
1254 	&	67	&	31	&	14.9	&	$-$17.17	  &	  -13.31  &	09	\\
1261 	&	50	&	42	&	22.5	&	$-$18.47	  &	  -12.53  &	42	\\
1304 	&	35	&	31	&	16.2	&	$-$16.86	  &	  -12.23  &	33	\\
1308 	&	39	&	27	&	11.4	&	$-$16.50	  &	  -11.32  &	44	\\
1333 	&	41	&	28	&	18.5	&	$-$15.44	  &	  -11.76  &	10	\\
1348 	&	42	&	25	&	13.1	&	$-$16.94	  &	  -12.83  &	23	\\
1353 	&	28	&	31	&	08.8	&	$-$15.51	  &	  -10.64  &	53	\\
1355 	&	24	&	28	&	29.6	&	$-$17.59	  &	  -10.96  &	57	\\
1389 	&	30	&	31	&	12.8	&	$-$15.98	  &	  -10.78  &	36	\\
1407 	&	27	&	31	&	11.8	&	$-$16.95	  &	  -10.99  &	52	\\
1661 	&	34	&	29	&	18.9	&	$-$16.18	  &	  -11.18  &	16	\\
1826 	&	23	&	31	&	07.8	&	$-$16.30	  &	  -11.91  &	56	\\
1861 	&	30	&	31	&	18.4	&	$-$17.78	  &	  -11.83  &	47	\\
1945 	&	31	&	38	&	21.5	&	$-$17.11	  &	  -11.66  &	35	\\
2019 	&	31	&	29	&	18.1	&	$-$17.53	  &	  -11.34  &	37	\\

\hline
 \end{tabular}
 The second and third columns are the measured signal-to-noise ratio (SNR) per pixel at 5000\AA\ for the
    galactic-light-subtracted nuclear spectra and the galactic
    main body spectra, respectively. The fourth column gives the
    half-light semi-major axis in SDSS {\it r} from Lisker07. The
    fifth and sixth columns are total galactic and 
    nucleus absolute magnitudes in SDSS {\it r},
    applying a distance modulus of $m-M=31.09$ mag \citep{Mei07},
    corresponding to $d=16.5$ Mpc. The last column represents the
    amount (in fraction of total light of the central aperture) of
    light subtracted from the central nucleus spectrum.  Nucleus magnitudes were derived as
    described in \citet{Paudel10b}: a two-dimensional elliptical
    model image of the galaxy, based on a S\'ersic fit to the
    radial profile, was subtracted from the original image, taking into
    account the median SDSS PSF of 1.4" FWHM. The nucleus magnitude
    was then measured by circular aperture photometry with $r = 2"$
    centered on the nucleus; the error is estimated to be 0.2 mag.

\end{minipage}
\end{table}

\begin{figure*}
\includegraphics[width=18cm]{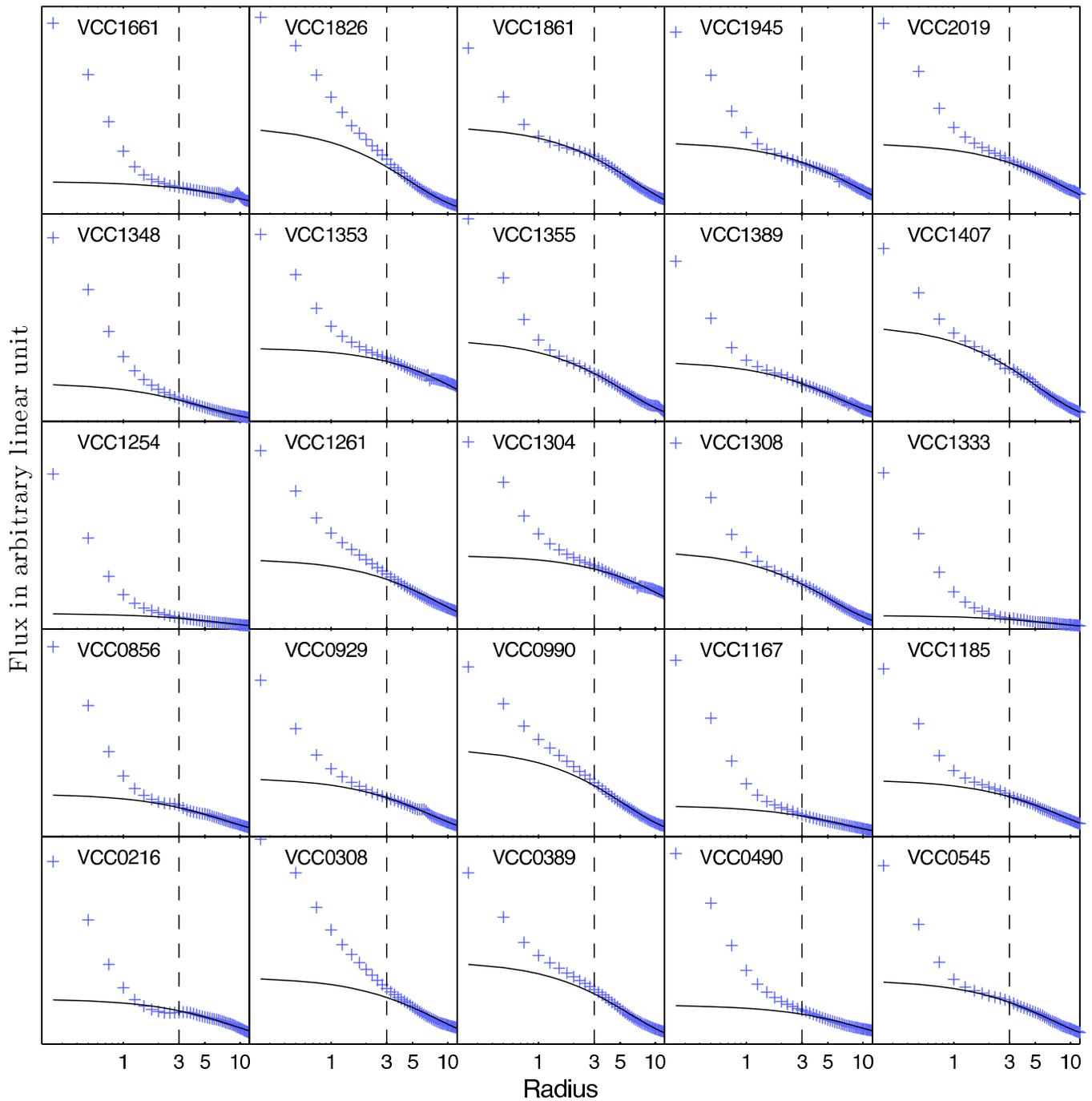}
 \caption{The light profile of dEs. The crosses represent the observed
   flux along the slit, and the solid line is the fitted exponential 
profile beyond 3" and extrapolated to the centre. }
 \label{fitser}
\end{figure*}

Our goals in this paper are the measurement of simple stellar
population (SSP) equivalent parameters (see \citealt{Trager08}) of dE nuclei and a comparison
with the SSPs of the surrounding galactic main bodies. Additionally, if the signal-to-noise ratio (hereafter SNR) 
permits us, we wish to explore gradients in the SSP parameters, which helps to determine whether the SSP 
of the nuclei is very different from the rest of the galaxy or is just a continuity of
a smooth SSP gradient at the centre of dEs. Although there is no precise
definition for what a nucleus is, the working definition used by several
studies is that an excess of light from the smooth exponential
(or higher order S\'ersic) profile of the rest of the galactic part is
observed, looking 
like a compact source sitting at the centre of the galaxy.
Because of its compactness, it is considered as a point source and
represented with a seeing convolved Gaussian light
profile. Likewise, the study of \citealt{Grant05} represents the
nuclei as a point source convolved with Gaussian seeing.
\citealt{Cote06} used a slightly different approach, by fitting a two
component core-S\'ersic model \citep{Graham03}. In Fig. \ref{fitser}
we can clearly see for most dEs the change in the light profile at the
centre (e.g.\ VCC0216, VCC0856, VCC0545, VCC1353 and VCC1945). On the
other hand, VCC0308, VCC0990, VCC1261 and VCC1826 exhibit a
rather smooth light profile. There may be several factors which
produce such differences in the light profile even though all dEs in this
sample are confirmed as nucleated from other photometric studies
\citep{Binggeli85,Lisker07}. Insufficient spatial
resolution or observed seeing which might blur the steeper light
profile of the nuclei makes it harder to separate the galactic light
profile. However, \citet{Cote06} have observed the existence of a profile
break in the case of VCC0856, VCC1261, VCC1355, VCC1407, VCC1661 and
VCC2019, reconfirming the existence of a nucleus at the centre of
dEs with HST high resolution surface photometry.

It is rather difficult to carry out an analysis of the stellar
populations of nuclei \emph{alone}, because the nuclei are always situated on
top of the underlying galactic main bodies. It is also hard to
separate the galactic light from the central nucleus of such a faint
object. The studies that have been done by \citet{Chilingarian09}
and \citet{Koleva09} provide results without galactic light
subtraction from the nucleus.  Although there is, in our sample, typically a fairly large
domination of light from the nucleus as compared to galactic light at
the photometric centre of the dEs, still a considerable amount of
underlying light of the host galaxy can alter the observed
properties of the nuclei. We therefore aim to reduce the
galactic light contamination in the nucleus spectra, attempting a
separate extraction of spectra for the nucleus and the galactic part.

We extract the nucleus spectra from the central 0".75 (i.e.,
3 pixels). The region between 0".75 and 3" is not used for this
extraction, to
avoid any effects of nucleus light in the spectra of the galactic main
body. We then integrate over the interval 3" to 8" from each side
of the nucleus to extract the spectra of the galactic main body (see Appendix A, Fig.~\ref{sef}). The
individual spectra of galactic main body from the different side of
nucleus were then co-added to produce a spectrum of higher SNR.
In order to subtract the galaxy light from the nucleus, we determine a
scaling factor by fitting the galactic main body's light profile
(measured along the slit) by an exponential profile and
extrapolating it to the very centre, yielding the amount of galaxy
light contained in the nucleus aperture (see Appendix A).
Given the above considerations
  about the difficulty of separating nucleus and galaxy, we point out
  that our approach ensures the removal of a \emph{significant part}, yet
  probably not 100\%  of galaxy light contamination. For those few cases
  where the central light profile looks rather smooth with
  ground-based data, and can only be disentangled with space-based
  photometry, our ``nucleus'' spectrum thus needs to be considered
  representative for the combination of nucleus \emph{and} galactic
  central light.

Before co-adding the spectra from the different sides of the galaxies,
we analyze their slit profile to check for inconsistencies or
asymmetries, e.g.\ by contaminating objects on the slit. We find only one
galaxy, VCC1945, has an asymmetric profile that deviates from a smooth
exponential profile on one side. We noticed that a bright point source
(foreground/ background or intra-galactic globular cluster) lies on one side of the slit. Therefore, we remove the 
spectrum from this
side. For completeness, we also compare the spectra from the 
different sides of the galaxy before co-adding them, and we
always find good agreement. Finally, the measured SNR
at 5000\AA{} for both the galaxy-subtracted nucleus spectra and
the combined galaxy spectra is given in Table \ref{snr}.

\section{Line strength measurements}

Before measuring the Lick absorption line indices from the flux calibrated
spectra of the galactic main bodies and nuclei, we also carefully
checked whether any emission lines are present, particularly since
some dEs show a fairly young nucleus. However, we do not detect any
[O{\small III}] emission, thus we do not correct the
H$\beta$ absorption for possible contamination by emission. If
such emission were present, it would make the measured H$\beta$
absorption smaller, and therefore derived ages older. On the other hand, it could be possible that
we do not see any emission lines because of the low spectral resolution. To quantify what strength of 
an emission line in a high-resolution spectrum \citep[i.e, model of][]{Vazdekis10} would
be smeared out in a low-resolution like ours, such that it is not recognized visually, we select 
a model spectrum of age 2 Gyr, and added an emission of H$\beta$. We then degrade the spectrum to the 
low-resolution of 11 \AA. We find that the added emission line could have an effect of up to 12\% on the 
measured absorption line strength, which reveals a relatively small effect on the age.

Note that we have not applied a velocity dispersion correction for the
Lick indices, because the expected galactic velocity dispersion,
$\sigma_{gal} \le 50$ km\,s$^{-1}$, is significantly below our
spectral resolution  $\sigma_{instr} \sim 280$ km\,s$^{-1}$. Therefore
these corrections are not necessary.

To measure the absorption line strengths from the spectra, we use the routine
{\it
  Indexf}\footnote{http://www.ucm.es/info/Astrof/software/indexf/indexf.html}
developed by N. Cardiel. It uses the definition of the Lick indices
from \citet{Trager98} and also derives the uncertainty in measured
strength using Monte-Carlo simulations. Calibrations of our
measured line strengths to the actual Lick system have been done as
 described in Paper I (Section 4.2 and Appendix B in that paper).  \\

We use the method of Lick indices \citep{Burstein84,Worthey94a,Trager98} as a
tool for estimating the stellar population characteristics.
We translate our Lick index measurements into SSP-equivalent ages, metallicities, and
$\alpha$-element abundance ratios by comparing them to the stellar population models of
\citet{Thomas03} by $\chi^{2}-$minimization, following
\citet{Proctor02}. For this we use the nine indices H$\delta_F$, H$\gamma_F$, Fe4383,
H$\beta$, Fe5015, Mg\,$b$ , Fe5270, Fe5335 \& Fe5406. Note that the SSP
models assume all the stars were formed in a single burst and have the
same age and metallicity. In fact, the galaxies may be a composite
stellar system formed during several episodic star formation events, with
different chemical compositions in general. Therefore, our estimated
stellar population parameters can be considered \emph{SSP$-$equivalent stellar
  populations}. The correlation of age and metallicity in the model fitting is illustrated in Appendix \ref{degen}.

\section{Results: Ages, metallicities and alpha-abundance ratios }

\begin{table*}
 \centering
 \begin{minipage}{150mm}
  \caption{SSP-equivalent stellar population parameters for the nuclei and the galactic main bodies.}
  \label{agmtt}
\begin{tabular}{l  |rr| rr rr|}
\hline
Galaxy Name & \multicolumn{2}{c}{Age, Gyr}  & \multicolumn{2}{c}{[Z/H], dex} & \multicolumn{2}{c}{[$\alpha$/Fe], dex}\\
\hline
     & Nuc. & Gal. & Nuc. & Gal. & Nuc. & Gal.\\
\hline
VCC0216	&	1.4	$^{	+0.3	}_{	-0.3	}$	&	4.0	$^{	+1.8	}_{	-1.2	}$	&	$-$0.61	$\pm$	0.15	&	$-$0.63	$\pm$	0.22	&		0.09	$\pm$	0.08	&	0.03	$\pm$	0.18	\\
VCC0308	&	1.5	$^{	+0.1	}_{	-0.1	}$	&	3.6	$^{	+1.6	}_{	-0.9	}$	&	 0.01	$\pm$	0.10	&	$-$0.34	$\pm$	0.17	&		0.42	$\pm$	0.09	&	$-$0.07	$\pm$	0.14	\\
VCC0389	&	4.1	$^{	+2.1	}_{	-1.3	}$	&	9.1	$^{	+3.4	}_{	-1.9	}$	&	$-$0.24	$\pm$	0.17	&	$-$0.43	$\pm$	0.20	&		0.17	$\pm$	0.12	&	$-$0.15	$\pm$	0.15	\\
VCC0490	&	1.9	$^{	+0.7	}_{	-0.2	}$	&	3.6	$^{	+2.1	}_{	-1.1	}$	&	$-$0.02	$\pm$	0.22	&	$-$0.24	$\pm$	0.17	&		$-$0.11	$\pm$	0.11	&	$-$0.11	$\pm$	0.15	\\
VCC0545	&	6.9	$^{	+2.6	}_{	-1.2	}$	&	12.5$^{	+0.0	}_{	-1.6	}$	&	$-$0.78	$\pm$	0.20	&	$-$0.88	$\pm$	0.10	&		0.24	$\pm$	0.18	&	$-$0.23	$\pm$	0.20	\\
VCC0725$^{a}$	&	5.5	$^{	+1.4	}_{	-1.7	}$	&	--	  -- 					 	&	$-$1.00	$\pm$	0.25	&	$-$-	--					&		0.16	$\pm$	0.38	&			--	--	\\
VCC0856	&	1.9	$^{	+0.2	}_{	-0.1	}$	&	15.0$^{	+0.0	}_{	-5.1	}$	&	0.03	$\pm$	0.10	&	$-$0.61	$\pm$	0.07	&		$-$0.14	$\pm$	0.06	&	0.03	$\pm$	0.16	\\
VCC0929	&	3.2	$^{	+0.5	}_{	-0.4	}$	&	3.8	$^{	+1.4	}_{	-0.6	}$	&	0.11	$\pm$	0.07	&	0.03	$\pm$	0.10	&		$-$0.16	$\pm$	0.05	&	0.15	$\pm$	0.07	\\
VCC0990	&	2.3	$^{	+0.9	}_{	-0.4	}$	&	5.5	$^{	+2.1	}_{	-1.1	}$	&	$-$0.19	$\pm$	0.15	&	$-$0.31	$\pm$	0.17	&		$-$0.30	$\pm$	0.04	&	$-$0.01	$\pm$	0.12	\\
VCC1167	&	15	$^{	+0.0	}_{	-0.0		}$	&	7.5	$^{	+7.5	}_{	-2.3	}$	&	$-$1.15$\pm$	0.05	&	$-$0.65	$\pm$	0.22	&		0.09	$\pm$	0.16	&	0.09	$\pm$	0.18	\\
VCC1185	&	11.9$^{	+0.6	}_{	-2.4	}$	&	12.5$^{	+1.2	}_{	-1.1	}$	&	$-$1.37	$\pm$	0.05	&	$-$0.68	$\pm$	0.10	&		$-$0.22	$\pm$	0.33	&	$-$0.01	$\pm$	0.22	\\
VCC1254	&	5.7	$^{	+1.2	}_{	-1.2	}$	&	15.0$^{	+0.0	}_{	-9.0	}$	&	$-$0.43	$\pm$	0.10	&	$-$0.48	$\pm$	0.32	&		0.05	$\pm$	0.07	&	$-$0.11	$\pm$	0.14	\\
VCC1261	&	1.8	$^{	+0.1	}_{	-0.0		}$	&	6.9	$^{	+2.2	}_{	-1.4	}$	&	0.18$\pm$	0.00	&	$-$0.46	$\pm$	0.15	&		$-$0.10	$\pm$	0.07	&	0.07	$\pm$	0.12	\\
VCC1304	&	8.6	$^{	+4.4	}_{	-2.1	}$	&	4.5	$^{	+0.7	}_{	-2.0	}$	&	$-$1.22	$\pm$	0.20	&	$-$0.56	$\pm$	0.27	&		$-$0.30	$\pm$	0.10	&	$-$0.22	$\pm$	0.19	\\
VCC1308	&	1.8	$^{	+0.3	}_{	-0.2	}$	&	15.0$^{	+0.0	}_{	-10.2	}$	&	+0.16	$\pm$	0.12	&	$-$0.70	$\pm$	0.42	&		0.09	$\pm$	0.09	&	0.11	$\pm$	0.18	\\
VCC1333	&	7.9	$^{	+5.8	}_{	-1.0	}$	&	1.0	$^{	+0.4	}_{	-0.0		}$	&$-$1.05$\pm$	0.20	&	$-$0.97	$\pm$	0.20	&		0.07	$\pm$	0.20	&	0.04	$\pm$	0.37	\\
VCC1348	&	10.9$^{	+2.2	}_{	-1.4	}$	&	15.0$^{	+0.0	}_{	-1.3	}$	&	$-$0.80	$\pm$	0.10	&	$-$0.53	$\pm$	0.07	&		0.45	$\pm$	0.14	&	0.50	$\pm$	0.04	\\
VCC1353	&	3.2	$^{	+1.2	}_{	-1.2	}$	&	4.1	$^{	+1.6	}_{	-1.4	}$	&	$-$1.02	$\pm$	0.25	&	$-$0.58	$\pm$	0.22	&		$-$0.26	$\pm$	0.32	&	0.38	$\pm$	0.18	\\
VCC1355	&	1.8	$^{	+2.3	}_{	-0.5	}$	&	3.2	$^{	+1.8	}_{	-0.6	}$	&	$-$0.48	$\pm$	0.39	&	$-$0.34	$\pm$	0.22	&		$-$0.08	$\pm$	0.30	&	$-$0.04	$\pm$	0.16	\\
VCC1389	&	13.1$^{	+1.9	}_{	-3.1	}$	&	11.9$^{	+0.0	}_{	-2.0	}$	&	$-$1.27	$\pm$	0.20	&	$-$0.85	$\pm$	0.10	&		0.07	$\pm$	0.30	&	0.19	$\pm$	0.19	\\
VCC1407	&	2.6	$^{	+1.3	}_{	-0.7	}$	&	14.3$^{	+0.7	}_{	-8.6	}$	&	$-$0.12	$\pm$	0.17	&	$-$0.73	$\pm$	0.34	&		0.07	$\pm$	0.14	&	0.11	$\pm$	0.16	\\
VCC1661	&	9.1	$^{	+5.9	}_{	-1.5	}$	&	6.6	$^{	+3.8	}_{	-1.1	}$	&	$-$0.95	$\pm$	0.15	&	$-$0.36	$\pm$	0.22	&		$-$0.26	$\pm$	0.14	&	$-$0.30	$\pm$	0.04	\\
VCC1826	&	1.7	$^{	+0.6	}_{	-0.2	}$	&	11.4$^{	+1.7	}_{	-2.3	}$	&	 +0.13	$\pm$	0.17	&	$-$0.90	$\pm$	0.15	&		$-$0.07	$\pm$	0.13	&	$-$0.10	$\pm$	0.19	\\
VCC1861	&	3.8	$^{	+2.2	}_{	-1.0	}$	&	4.1	$^{	+1.3	}_{	-1.3	}$	&	$-$0.29	$\pm$	0.17	&	$-$0.12	$\pm$	0.12	&		$-$0.16	$\pm$	0.14	&	0.07	$\pm$	0.09	\\
VCC1945	&	6.6	$^{	+8.4	}_{	-1.3	}$	&	14.3$^{	+0.7	}_{	-2.4	}$	&	$-$0.75	$\pm$	0.27	&	$-$1.00	$\pm$	0.10	&		0.00	$\pm$	0.24	&	$-$0.30	$\pm$	0.23	\\
VCC2019	&	1.7	$^{	+0.2	}_{	-0.3	}$	&	8.3	$^{	+6.1	}_{	-2.5	}$	&	+0.06	$\pm$	0.15	&	$-$0.41	$\pm$	0.24	&		$-$0.27	$\pm$	0.12	&	0.00	$\pm$	0.16	\\
\hline
 \end{tabular}
$^{a}$without subtraction of galactic light and does not have a
measurement of SSPs from the galactic main body (see text).  
\end{minipage}
\end{table*}

\begin{figure*}
\includegraphics[width=15cm]{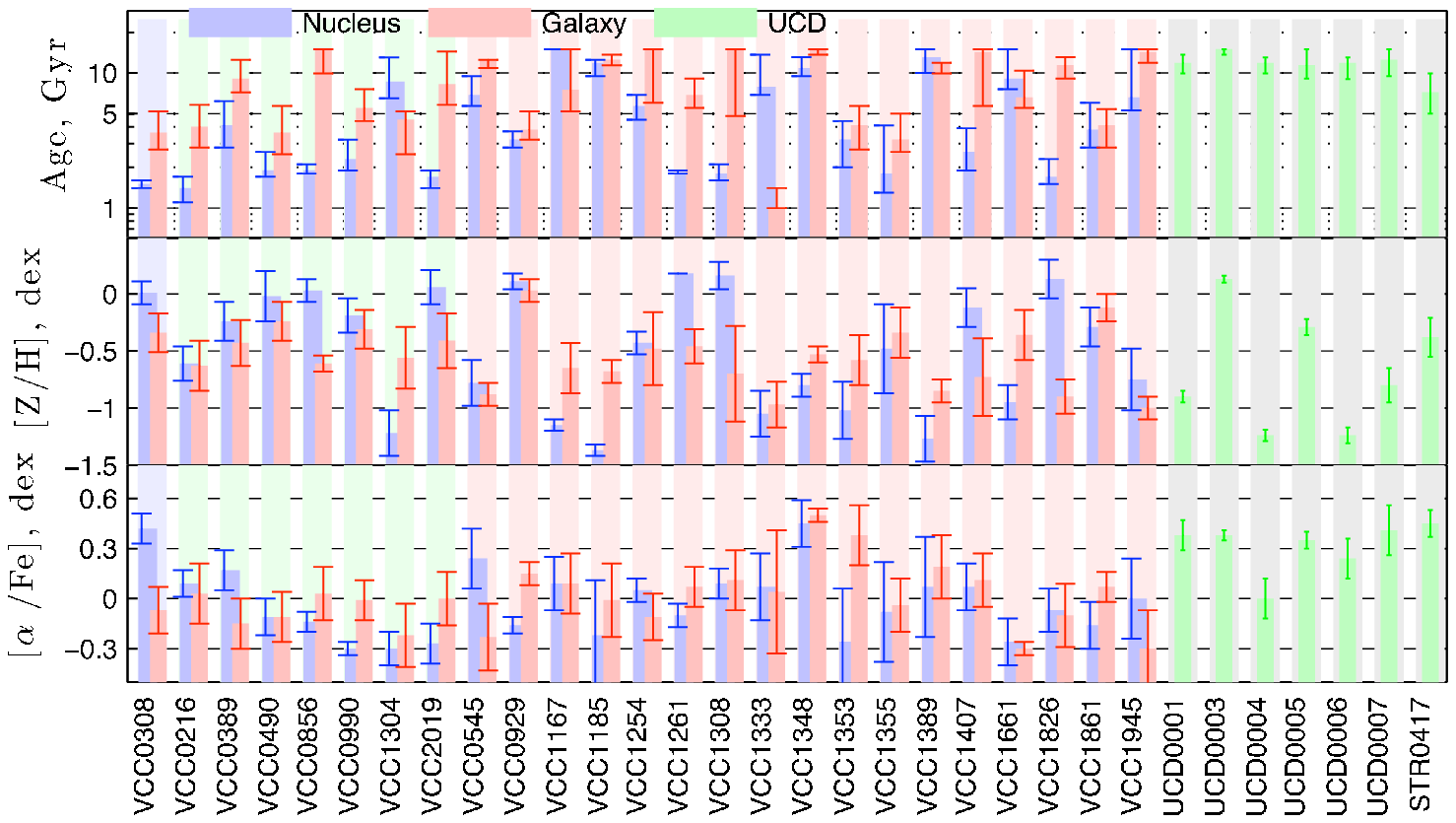}
 \caption{A comparison of stellar population parameters.  The SSPs from the different parts of the dEs are 
represented with vertical bars of different color: blue for the nuclei and
red for the galactic main bodies. The faint background 
colors indicate the dE subtype: blue for the nucleated dE with disk and blue centre,
green for the nucleated dEs with disks, and red for the nucleated dEs
without disk features. The Virgo UCDs are represented by the 
green vertical bars with gray background. For the UCDs, we used
published values of line strengths from \citet{Evstigneeva07} to derive  
the stellar population parameters (see text).}
 \label{alfd}
\end{figure*}

In this section, we present the SSP-equivalent ages, metallicities and
$\alpha$-abundance ratios of our sample dEs (Table \ref{agmtt}). Note 
that, in case of the least luminous dE, VCC0725, we find that the sky
noise becomes dominant beyond the central aperture. Hence, we remove
its galactic part from the sample and therefore provide no 
SSP parameters for the galactic main body of this dE.

We can clearly see that the ages of the nuclei are significantly
lower than the ages of the surrounding galactic main bodies
(Fig. \ref{alfd}). The differences are more prominent in  
the disky dEs: only VCC1304 has a nucleus that is older than the galactic part. Moreover,
we find that only four other non-disky dE(N)s (VCC1167, VCC1333,
VCC1389 and VCC1661 $-$ see Sec.\ref{sample}) have nuclei with
significantly larger ages than the galactic main bodies. The median difference
in age between the galactic main
bodies and nuclei is 3.5 Gyr. Examining Fig.
\ref{alfd} individually galaxy by galaxy, one can see that VCC0856
shows the largest difference ($>$10 Gyr) in age between the nucleus
and the galactic part. The nucleus of the  blue centre dE
VCC0308, while having a young age, does not show up as being special,
having an age of
1.5 $\pm$0.1 Gyr, similar to other dE nuclei such as VCC0216,
VCC2019 and VCC1826.

The metallicity distributions of the nuclei and the surrounding galactic
main bodies also differ: the majority of the
nuclei are relatively metal enhanced as compared to the galactic main
bodies. However, it is remarkable that those nuclei that are older
or equally old as the galactic part are also less metal rich than the
latter. We find that the nucleus of VCC1308 has the highest
metallicity of +0.16 $\pm$0.12 dex. For all dEs, the galactic
main bodies have sub-solar metallicity. The $\alpha$-abundance
ratio from nuclei and galactic main bodies show a wide distribution. The nuclei of three dEs (i.e.,
VCC0308, VCC0389 and VCC0545) show significant $\alpha$-enhancement as
compared to their galactic part.  On the contrary, 
four dEs, VCC0990, VCC0929, VCC1353 and VCC1861, exhibit a significantly
enhanced $\alpha$-abundance in the galactic part as compared to their
nucleus.

In the right part of Figure \ref{alfd}, the green vertical bars
present, for comparison, the derived stellar population parameters of
the UCD sample of \citet{Evstigneeva07}. Note
that we only use the published four indices (H$\beta$, Mg{\it b},
Fe5270 and Fe5335). However, we use the same method of
estimation for the stellar population parameters. The UCD ages and metallicities are
consistent with old and metal poor stellar populations. Almost all
UCDs have ages  $\sim$10 Gyr and metallicities vary between -1.25 to 0.13 dex.
The [$\alpha$/Fe]-abundances  are always super solar in
case of the UCDs, with a mean of 0.31 dex,  which is 0.34 dex higher than the mean
[$\alpha$/Fe] of the dE nuclei.

\begin{figure}
\includegraphics[width=8.5cm]{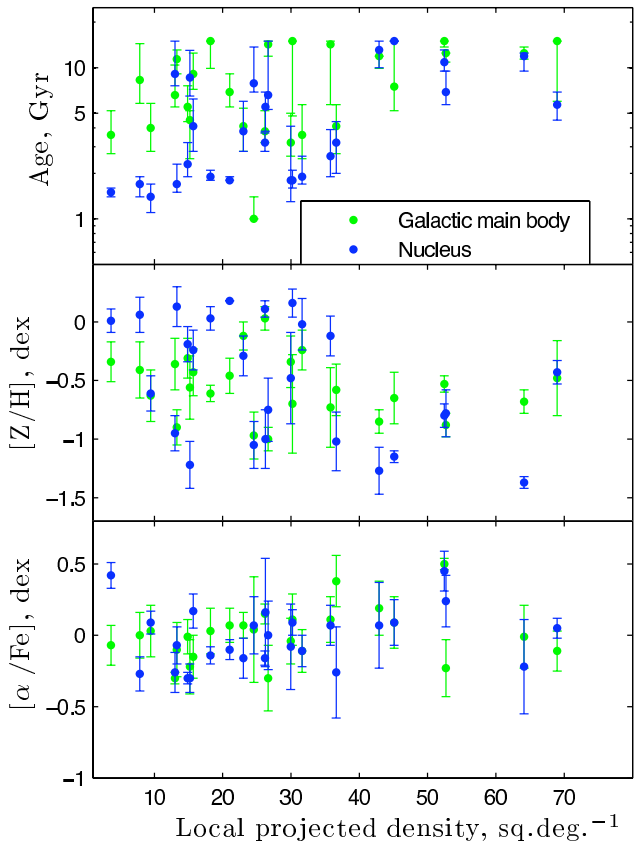}
 \caption{The age, metallicity and [$\alpha$/Fe] versus local
   projected density. Green color represents the galactic 
main body and blue indicates the nucleus. }
 \label{env2}
\end{figure}

The relation between the stellar population parameters and the local
projected number density of galaxies in the cluster is plotted in Figure \ref{env2}. The local projected
density has been calculated from a circular projected area enclosing
the 10th neighbor. It seems that there is a correlation between
the local projected density and the ages of the nuclei. The Spearman rank order test shows a weak correlation
of the ages and metallicities of the dE nuclei with the local projected densities. The correlation 
coefficients are 0.5 and $-$0.4, and the probabilities of the null 
hypothesis that there is no correlation are 0.2\% and 4\% for the age and metallicity, respectively. 
Unlike this, a similar test shows that the SSPs of the galactic main body do not have any 
relation with local projected densities.

 \begin{figure}
\includegraphics[width=9cm]{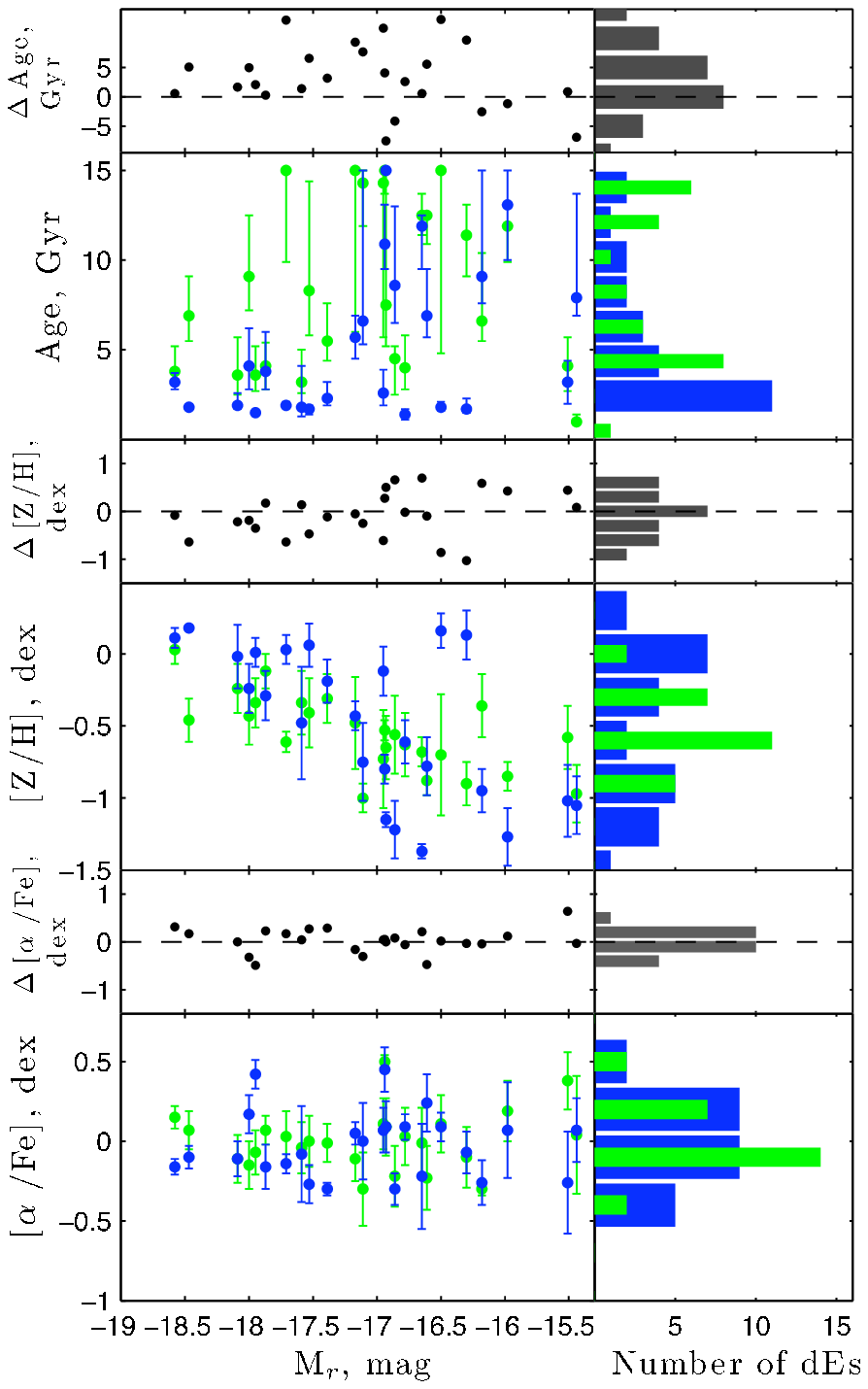}
 \caption{The derived ages (top), metallicities (middle) and
  [$\alpha$/Fe]-abundance (bottom), plotted against $r-$band absolute magnitude (left). The blue color represents 
the nuclei and green color indicates the galactic main body. On top of each panel, we also show the difference in 
the SSP parameters, i.e. galactic part $-$ nucleus. In the right
panel, we provide the number distribution of the parameters. } 
 \label{env}
\end{figure}

The relations between the stellar population parameters and the
total galactic luminosity are presented in Fig. \ref{env}. At the
top of each panel, we also provide the trend of the differences in the SSP
parameters between the galactic main bodies and the nuclei. It is
clearly recognized that almost all dEs brighter than $M_{\rm r}=-17$ mag have
younger and more metal-rich nuclei than the galactic main bodies. On the
other hand, 
there is a relatively large scatter in the low luminosity region, and
we can see that some of the nuclei are as old  and metal poor as the galactic main bodies.
However, the sign of the differences in age and metallicity between
galactic main body and nucleus are completely opposite at the fainter
and brighter end of the plot. As there exists a well-known
metallicity-luminosity relation in early type galaxies
\citep{Poggianti01}, our sample also follows this relation for both 
nuclei and galaxies, i.e., the  metallicity decreases with
decreasing total galactic luminosity.  The derived [$\alpha$/Fe] values are fairly consistent with
a roughly solar value for both nuclei and galactic main bodies.

In the right panels of Fig. \ref{env}, we provide the number
distribution (in the histogram) of stellar population
parameters of the nuclei (in blue color) and the galactic main bodies
(in green color). It seems that the ages of the galactic main
bodies have a bimodal distribution, but the small number of data points
in each bin and the fairly large errors in the age measurement increase
the uncertainty; the bimodality thus remains a qualitative
impression. The age distribution of the nuclei is highly
dominated by nuclei of younger ages. The metallicity distribution
however appears much broader in case of nuclei than galactic main
body. The nucleus metallicity ranges from slightly super-solar (+0.18
dex) to strongly sub-solar values (-1.22 dex), and interestingly all dE
galactic main bodies have sub-solar metallicity.

\subsection{Stellar population gradients}

\begin{figure}
\includegraphics[width=8.5cm]{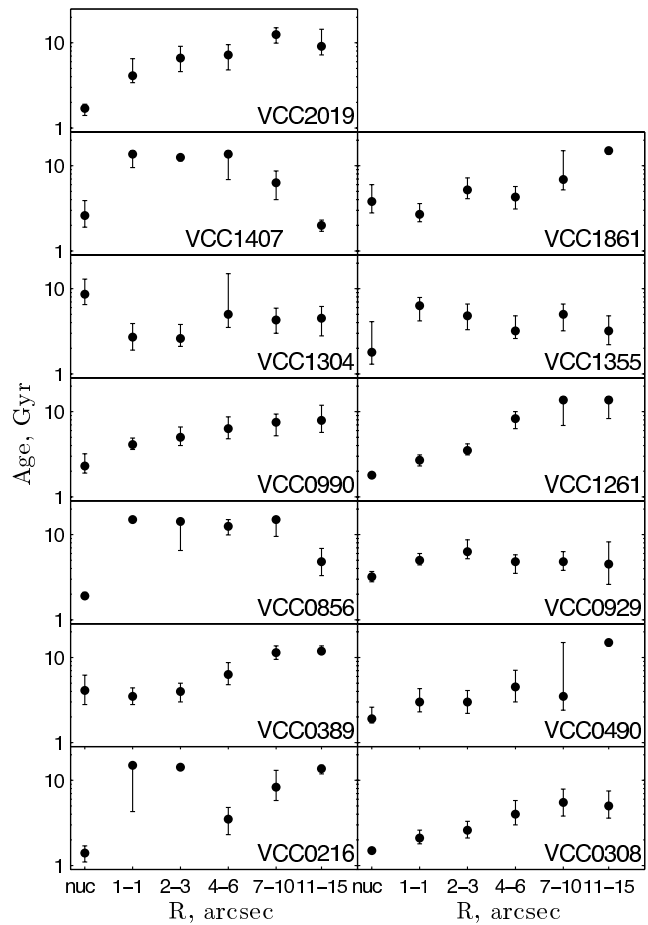}
 \caption{The radial age profiles of selected dEs (here we select those dEs which have sufficient SNR at the last 
radial bin, 11" to 15").  }
 \label{grag}
\end{figure}
\begin{figure}
\includegraphics[width=8.5cm]{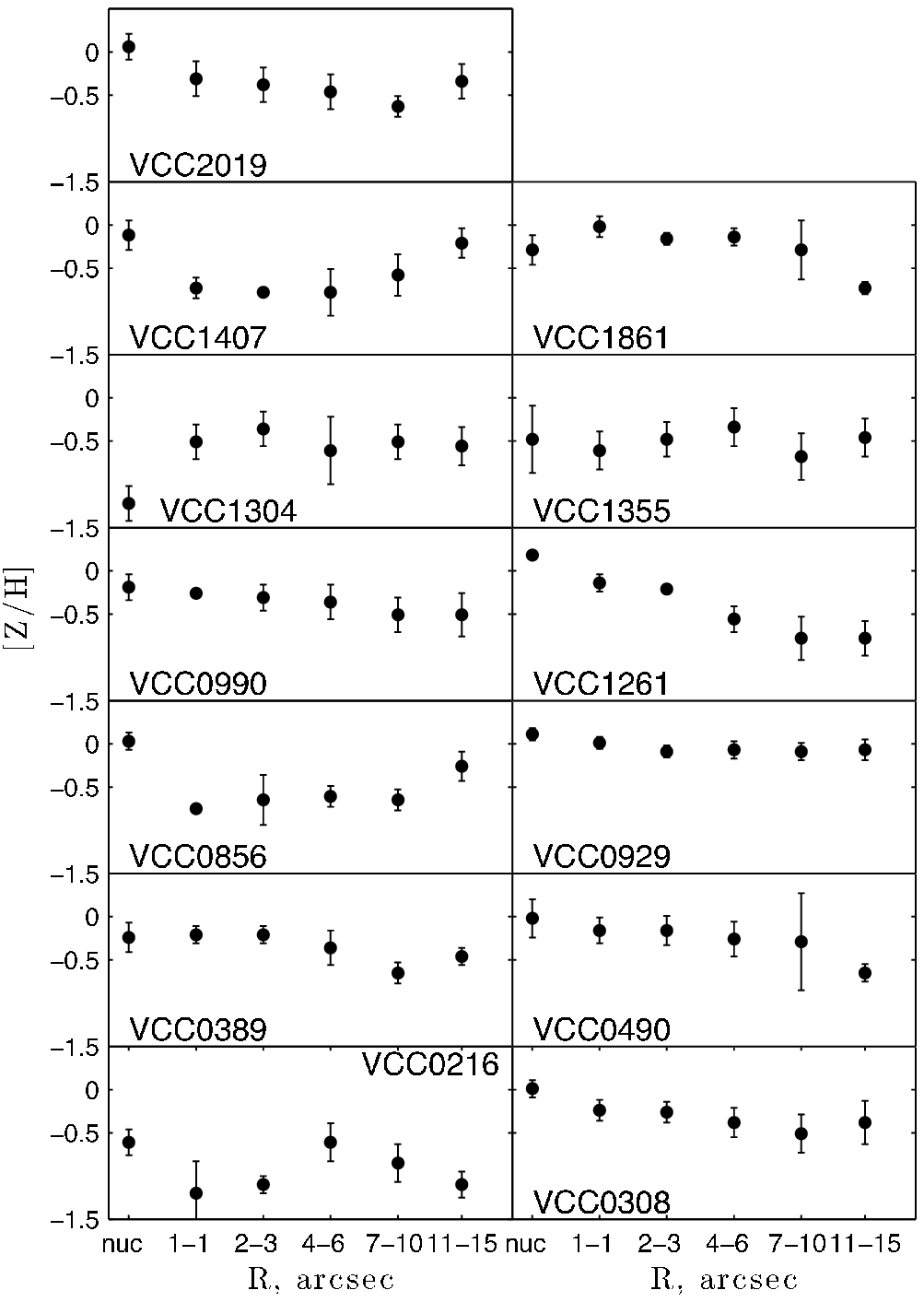}
 \caption{The radial metallicity profiles of the dEs, selected as in Fig. \ref{grag}.}
 \label{grmet}
\end{figure}

Due to the low brightness of dEs, it is always challenging to get spectra
from their outer part with sufficient SNR to study stellar
population gradients.  Some attempts have been made to derive the stellar
population gradients in the different cluster dEs
(\citealt{Chilingarian09} for Virgo, \citealt{Koleva09} for
Fornax). These studies used different methods to obtain SSP
parameters, namely through spectral fitting with SSP models. 
\citet{Chilingarian09} observed either flat or negative radial gradients in
metallicity in his sample. However, due to the relatively high
uncertainty in the age estimation, he did not draw conclusions on the
radial behavior of ages. The study of \citet{Koleva09}
reconfirmed the result of the existence of negative metallicity
gradients and found radial age gradients in the dEs, with older ages
at larger radii.

\begin{figure}

\includegraphics[width=8.5cm]{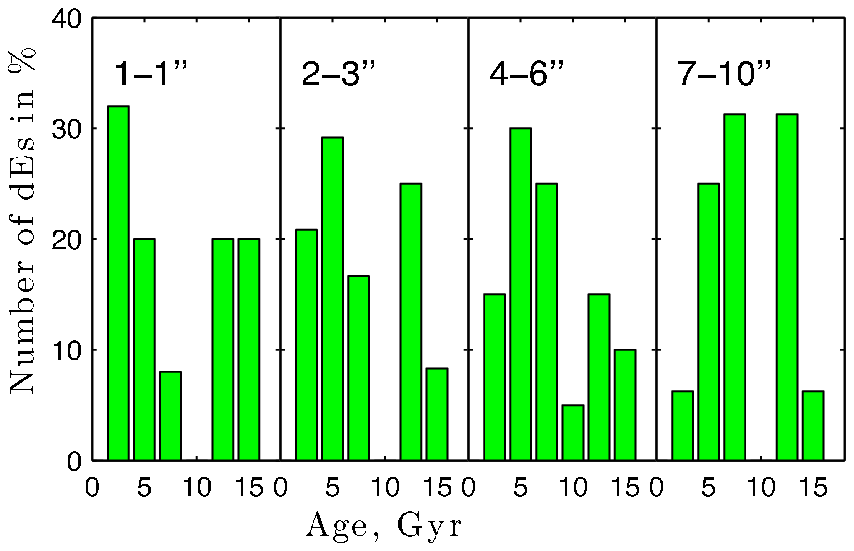}
\includegraphics[width=8.5cm]{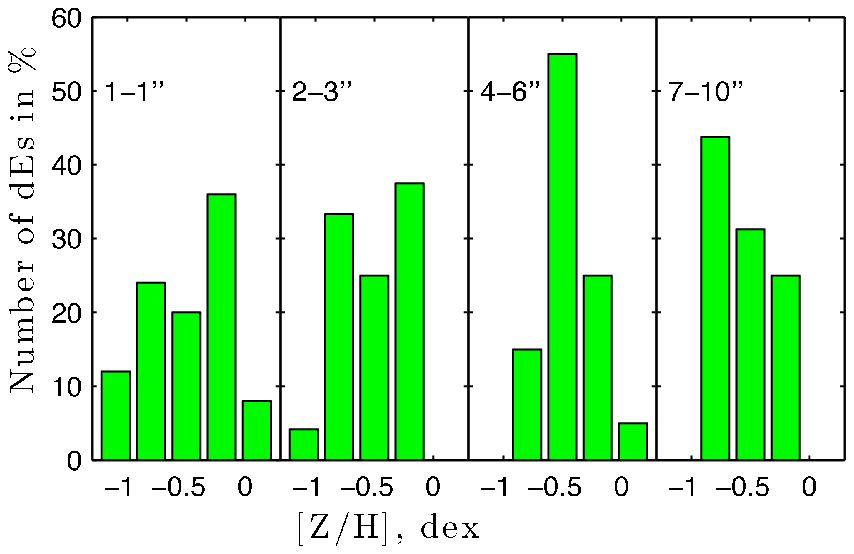}
 \caption{ The age and metallicity distribution at different radial bins. }
 \label{bhis}
\end{figure}

\begin{figure}
\includegraphics[width=9cm]{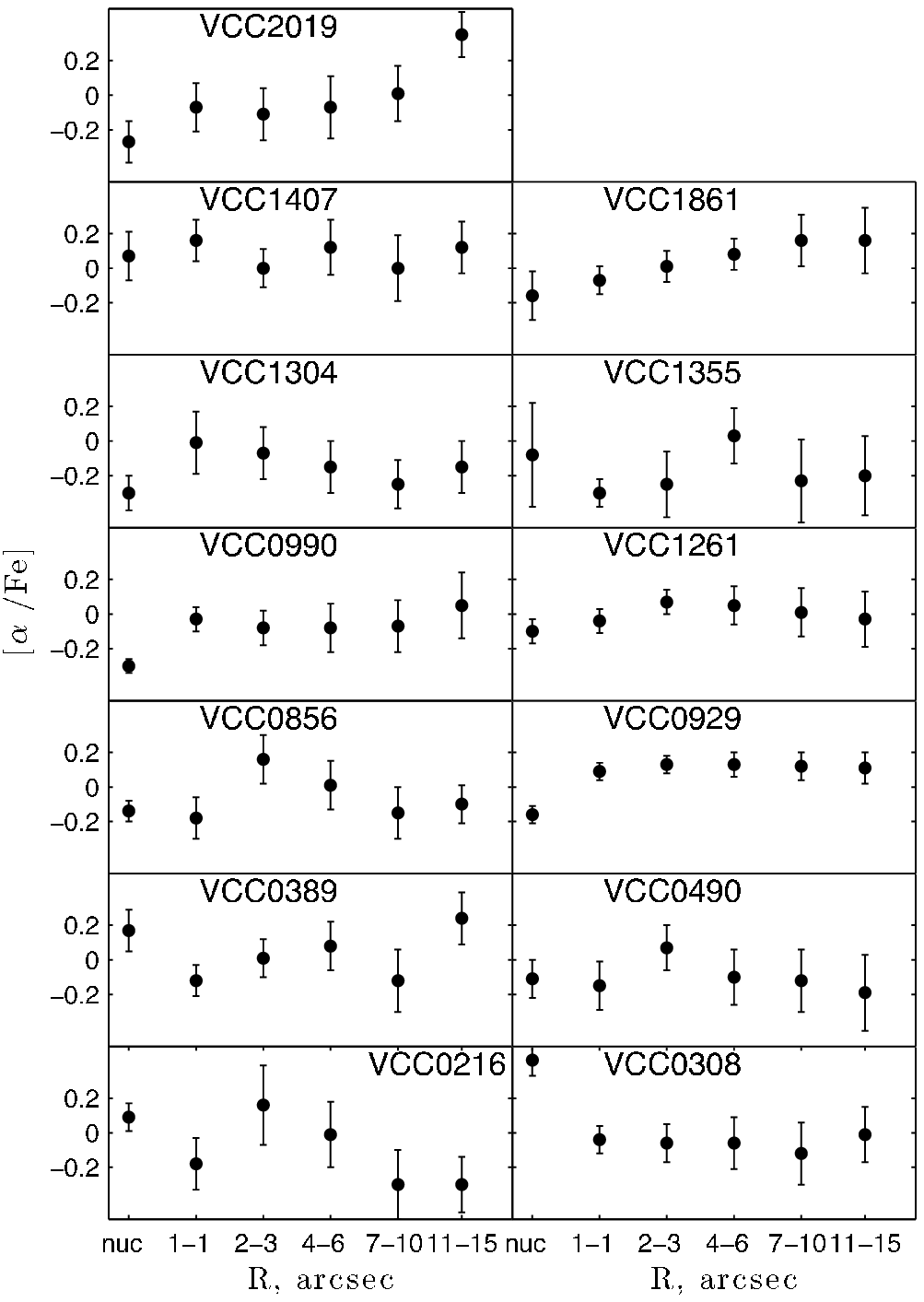}
 \caption{The radial profile of the $\alpha$-abundance ratios of the dEs, selected as in Fig. \ref{grag}.}
 \label{gral}
\end{figure}

In Figure  \ref{grag}$-$\ref{gral}, we present the radial profiles of
SSP-equivalent age, metallicity and abundance ratio, measured in bins
along the major axis of the dEs. It is interesting that we
can divide these trends of SSPs in two groups. The first group are those
dEs which exhibit a smooth trend of increasing the age and decreasing
metallicity with radius, beginning from the nucleus, such as
VCC0308, VCC0490, VCC0929, VCC1261 and VCC2019.  In contrast, the second
group shows a break in the SSP profile when going from the nucleus to
the surrounding galactic part, with the latter having a nearly flat
gradient, like e.g.\ for VCC0216, VCC0856, VCC1304 and VCC1355. 

Three dEs, VCC2019, VCC1261 and VCC0308, show a significant gradient in age
and metallicity, having a relatively young and metal enhanced
nucleus. Likewise, the ages of VCC0389, VCC0490, VCC0990,
VCC0929 and VCC1407 also seem to correlate with the radius. Our
derived ages
for VCC0856 agree with the result of \citet{Chilingarian09} that this
galaxy has a flat distribution of ages beyond the central nucleus. In addition to
that, we can also see such a flatness in the age distribution of
VCC1355. VCC1261 presents the
largest gradient in metallicity starting from slightly super solar
down to a sub-solar value of $-$0.75 dex. Although we do not see any strong
trend of [$\alpha$/Fe] with radius in most of the cases, VCC0216
and VCC2019 display the opposite trend of decreasing and increasing of
[$\alpha$/Fe] with radius, respectively. 

In Figure~\ref{bhis} we show the age and metallicity distribution of
our dEs in the different radial bins.
Note that there is not always the same number of dEs in each radial
bin: due to insufficient SNR in the outer radii for some dEs, those
were omitted from the respective bins.
 The first  1" bin contains 25 dEs, and the second, third and fourth
 bin contains  24, 20 and 16 dEs, respectively. Therefore, the y-axis
 represents the normalized fraction in 
percent. It is easily noticeable that the distributions change with
radius: the inner bin is dominated by young ages and shows a broader
metallicity distribution, and the fraction of old ages and low
metallicities increases as we go outward, with the metallicity
distribution becoming narrower.

\section{Discussion}

In this paper, we have characterized the stellar population parameters
from the different parts of dEs: the nuclei and the surrounding
galactic main bodies. Our primary
motivation for this is to improve our understanding of the physical mechanisms responsible for
the formation of dE nuclei and the subsequent evolution of dEs
themselves. As we now discuss, our study makes two important contributions
in this context: (i) to much more firmly establish the SSP-equivalent
stellar population parameters of dEs and their nuclei (ii) to cast new light on
the spatially resolved stellar population characteristics of dEs.

The surrounding galactic main body is represented by extracting its
spectrum from a 5" radial interval beyond 3" from the centre, avoiding any
contamination with light from the nucleus. 
We expect that, due to our method of subtraction of the underlying
galactic light from the nucleus spectra (see
Appendix \ref{lsub}), we obtained comparatively clean spectra of the
nuclei, with the derived stellar population properties
from such spectra well representing the nucleus stellar
population.
 Nevertheless, as outlined before, in the cases
of weak nuclei there is still a chance that the remaining galactic
light contributes significantly, such that the nuclear spectra
represent the combination of nucleus and ``central
galaxy light''.
To test for a possible bias due to this effect, we select those dEs which
have galactic light fraction (see Table \ref{snr}) larger than 50\% at the central
aperture such as VCC0990, VCC1353, VCC1355, VCC1407 and VCC1826, but
all these nuclei have ages less than 5 Gyr, and agree
fairly well with the average age of the nuclei in total.

Generally speaking, stellar population gradients can be used as a proxy for
the study of the evolutionary history of early type
galaxies, since different formation models predict different
gradients. In a nutshell, monolithic collapse models \citep{Arimoto87}
predict slightly steeper gradients than the hierarchical merging model
\citep{White80}. These predictions, however, mainly apply to
normal early-type galaxies (Es). In case of early-type dwarfs, different
formation scenarios might be relevant, such as morphological transformation, or
simply a primordial origin (also see the discussion in Paper
I). Nevertheless, the overall distribution of  
age and metallicity at the
different radial bins suggest that it occurs more frequently that the inner
parts of dEs are younger and more metal enhanced than their outer
parts, which is consistent with previous studies
\citep{Chilingarian09,Koleva09}. We also see two distict behaviours of
radial SSP profiles; the
presence of flat profiles may be due to a particular galaxy structure
(i.e., a faint underlying disk) or may be an indication of a 
different origin.
Among the dEs with smooth SSP gradients, VCC0308 only has a very weak blue
centre \citep{Lisker06b}, so it may well be that other galaxies have just a bit
weaker colour gradients and were thus not labeled ``blue-centre dE'' previously. On
the other hand,  VCC0216  and VCC0856 have a similarly young
nucleus as VCC0308, but not an age gradient in the galaxy itself, 
which might  lead to having no colour gradient.

Another key result emerging from our study is a very clear picture of the differences between the stellar 
populations of the nuclei and the galactic main
bodies of the dEs. To our knowledge, no spectroscopic study has yet performed
such a comparison with a similar sample size.
Studies based on color differences (\citealt{Durrell97}, \citealt{Cote06}, and
particularly \citealt{Lotz04b}) find slightly bluer nuclei. It is, however,
not straightforward to interpret these color differences in the
sense of stellar population properties, as we know that a
degeneracy in the age and metallicity exists with color (see also Appendix \ref{degen}). In
contrast to the explanation of \citet{Lotz04b} of having more metal rich
populations in the surrounding galactic main bodies, we find a metal poorer and older
population in the galactic part on average. In addition to this, as
\citet{Cote06} note, there exists a color-luminosity relation
for the nuclei. We also find that the metallicity of dE nuclei correlates
with the total luminosity of dEs. 

We
have seen that there is almost no correlation between the ages of
the galactic main bodies and the luminosity of the dEs. This might, at
first glance, imply that the reason
for the apparent age dichotomy in Paper I, finding a clear correlation
with luminosity for the central stellar populations of dEs, 
was due to the nucleus contribution to the central aperture light. However, Fig. 
\ref{diff} of the Appendix, which compares 
the SSPs resulting from the
nucleus spectra before and after subtraction of the underlying
galactic light, actually tells us that this conclusion is not true: if
the very central stellar populations of the galaxies, whose pure light cannot
be seen due to the superposed nucleus, would be so much older than
the nucleus itself, the difference
before/after subtraction would be quite significant, which is not found.
Instead, the figure tells us that the very central part of the galaxy
does also reach, in most cases, almost the young age of the
nuclei. Thus, in many cases
it is really the age gradient within the galaxy that makes the galactic
part surrounding the nucleus appear significantly older than the
nucleus itself in Fig. \ref{env}.

\subsection{Evolution of dEs and formation of nuclei}
As we mentioned in the introduction, many studies have discussed the
origin of the nuclei of dEs together with the evolution of dEs
themselves. It is challenging to provide definitive observational
tests of these different scenarios. Moreover, we argue in Paper I that
not all dEs are the same class of object. The dichotomy  
in the age distribution 
of the galactic main bodies also supports the idea 
that one type of dEs may have a primordial origin \citep{Rakos04},
being relatively old and metal poor. These might
have suffered either early infall into the cluster potential or formed
together with the cluster itself. 
The common idea is that internal feedback might be responsible for the
removal of gas, with the consequence that star formation activity
ceases at such early epochs.  

On the other hand, dEs with a relatively young and metal enhanced 
galactic main body likely have a different origin. As they are also
preferentially brighter and often host
disk-structure, they might have formed through the structural transformation of a late-type spiral 
into a spheroidal system, triggered by the popular scenario of strong tidal interactions with massive cluster 
galaxies. Simulations have shown that late-type galaxies entering in a rich cluster can undergo a significant 
morphological transformation into spheroidals by encounters with
brighter galaxies and with the cluster's tidal field  
\citep{Moore96, Mastropietro05}. This scenario is unlikely to produce the observed 
radial SSP gradients: either metallicity gradients must have formed in the late-type 
galaxies and somehow preserved during morphological transformation
\cite[see the discussion in][]{Spolaor10a}, or accretion 
of leftover gas towards the centre of the galaxy would have to be responsible
for the creation of such gradients. However, the  flat [$\alpha$/Fe]
profile implies a similar star formation time scale everywhere in the
dEs.

As we discussed above, the fairly different types of dEs with and without disk structure might have a different
origin. It is therefore even more difficult to explain the origin of the nuclei of these dEs
with a single scenario. However, from this and previous studies, it is
becoming clear that the majority of dE nuclei
are unlikely to have formed through the merging of globular
clusters: \citet{Cote06} already explained the difficulty of this scenario
with the luminosity differences, and additionally we find that most
nuclei are fairly young and metal rich, at least in case of the brighter dEs (M$_{r}
\le-$17.25 mag). There are still the nuclei of some fainter
dEs (i.e., M$_{r}$ $>$ $-$17.25 mag) which have fairly old
and metal-poor populations, more resembling the stellar population properties
of globular clusters. They might have formed through a different
process as the nuclei of brighter dEs.

The younger and comparably metal-rich nuclei support the idea that the
central stellar populations of dEs were governed by continuous infall
and accretion of gas in the
centre of the potential well, building the nuclei. The brighter dEs
also host disk features (e.g. residual spiral arms/bars) and these dEs
themselves might have been formed through the transformation of
late-type spirals (Sc-Sd types). High resolution HST imaging has shown that such late-type objects frequently
contain a compact nuclear cluster \citep{Boker02,Boker04}, and \citet{Cote06} observed that such nuclear 
clusters have similar sizes to dE nuclei. Stellar population studies have shown that the 
majority of nuclear clusters have ages of few tens of Myr \citep{Seth06,Walcher06} with episodic star 
formation activity. Following the simplest interpretation, it could be that the present day
dE nuclei are simply the nuclear clusters of the transformed late-type galaxies, and their star formation activity 
faded with the morphological transformation of the host galaxies. However, this scenario again fails to explain the 
observed age difference between the nuclei and galactic main bodies, since late type disks are also considered to host 
star formation activity throughout the inner region and disk. Alternatively, the truncation of star formation in the disk due to 
interactions could be more efficient than in the nucleus, which eventually leads to the development of age/metallicity 
gradients in dEs and makes the 
central nucleus younger and metal richer than the galactic main body. In any case, more detailed
numerical simulations are required to test these hypotheses.


We find that dE nuclei exhibit fairly different  
stellar populations than UCDs. Particularly, the relatively older population (larger than 8 Gyr) and 
slightly super-solar $\alpha$-abundance of UCDs may seem to create an inconsistency in the idea of dE 
nuclei being the progenitors of UCDs. Nevertheless, the current sample of UCDs is limited, and the fairly 
large spread in the stellar population properties of dE nuclei may allow the possibility of UCD formation in the 
Virgo cluster by the stripping of such dEs whose nuclei have old and
metal poor stellar populations \citep{Paudel10b}. Therefore, 
a larger sample of UCDs and perhaps a more rigorous comparison of SSP properties than this work is 
needed before any strong conclusions can be drawn.

\section{Conclusions}

We have investigated the stellar population properties of the central nucleus and the 
surrounding galactic main body for a sample of 26 dEs in the Virgo
cluster and compared the SSP-equivalent stellar population  
parameters of the dE nuclei with the ones of a small sample of
UCDs. In addition to this, we have derived the  radial profiles for
 age, metallicity and  
[$\alpha$/Fe] abundance for 13 dEs. Our main findings can be 
summarized as follows:

\begin{itemize}
\item We find that for most of the dEs the nuclei are significantly younger ($\sim$3.5 Gyr) and more metal rich 
($\sim$0.07 dex) as compared to the galactic main body of the galaxies. Only five dEs have significantly older 
nuclei than their galactic main bodies, and dEs with old and metal poor nuclei are more likely to be 
distributed in the dense region of the cluster than the dEs with young and 
metal-enhanced nuclei.

\item The metallicity of dE nuclei correlates 
with the total luminosity of dEs, and the observed metallicities of
the nuclei have a fairly large range (+0.18 to -1.22
dex). All galactic main bodies of the dEs have sub-solar
metallicity.


\item While we see two distinct behaviours of SSP profiles (with
  and without a break) the overall trend of  
increasing age and decreasing metallicity with the radius is consistent
with earlier studies. The $\alpha$-abundance as
function of radius is consistent with no gradient.

\item These observed properties suggest that the merging of globular clusters might not be the
appropriate scenario for the formation of nuclei in dEs, at least not
for the brighter dEs. The younger and comparably metal-rich nuclei support the idea that the
central stellar populations of dEs were governed by continuous infall/accretion of gas in the
centre of the potential well, building the nuclei.

\item The heterogeneous nature of the stellar population characteristics of
  dEs hints at different formation scenarios of
  dEs, similar to the conclusion of our previous study
  \citep{Paudel10}. Our results suggest that the old, faint and
  metal-poor dEs are more likely to have 
a primordial origin, while those with relatively young ages and a higher
metallicity and luminosity may have formed through morphological transformation.

\end{itemize}

\section{Acknowledgments}
We thank Michael Hilker for useful comments. We thank the referee for
providing useful suggestions for improving the manuscript.
   S.P.\ and T.L.\ are supported within the framework of the Excellence
    Initiative by the German Research Foundation (DFG) through the Heidelberg
    Graduate School of Fundamental Physics (grant number GSC 129/1).  S.P. acknowledges 
the support of the
International Max Planck Research School (IMPRS) for Astronomy and
Cosmic Physics at the University of Heidelberg.
This work was based on observations made with ESO telescopes at
Paranal Observatory under programme ID
078.B-0178(A).
 This work has made
use of the NASA Astrophysics Data System and the NASA/IPAC
Extragalactic Data base (NED) which is operated by the Jet Propulsion
Laboratory, California Institute of Technology, under contract with
the National Aeronautics and Space Administration.

\appendix

\section{Subtraction of galactic light from the nucleus}\label{lsub}
\begin{figure}
 \includegraphics[width=8cm,height=9cm]{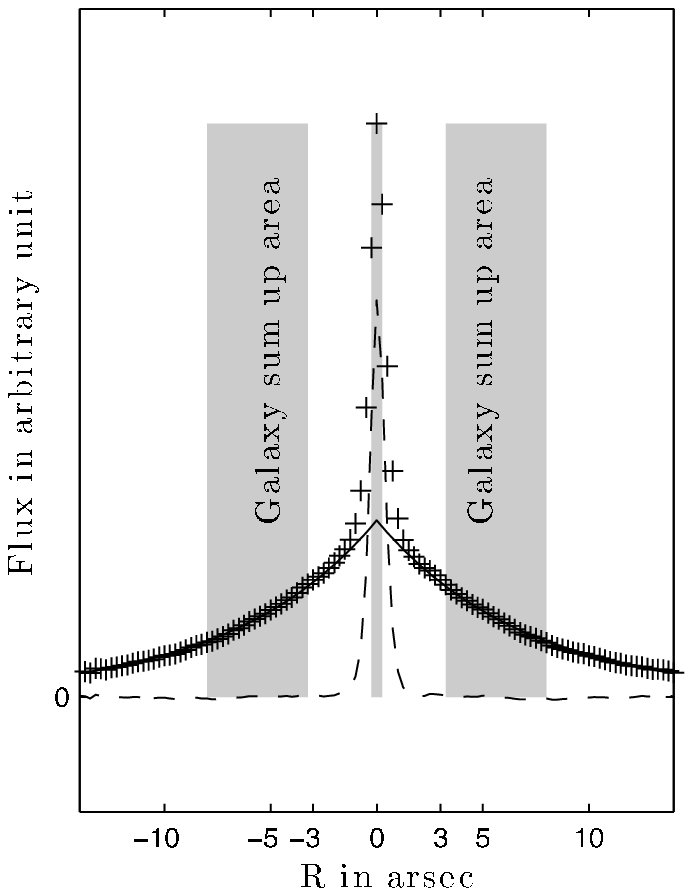}
 \caption{A schematic view of the fitting of the light profile for
   VCC0490 and the binning processes. The cross symbol  
represents the distribution of the observed total light (i.e galaxy + nucleus) and solid line represents the 
exponentially fitted 
light profile of the galaxy. The dashed line is the residual nucleus after the subtraction of galaxy light, which 
represents the pure nuclear light profile.}
 \label{sef}
\end{figure}

The Figure \ref{sef} provides a schematic view of the subtraction of
galactic light from the central nucleus.  
First, we average the galaxy frame in the wavelength direction between
4000 \AA{} to 5500 \AA, assuming that there is  
not any severe change in light profile from exponential with the
wavelength. The fitting of the galaxy light profile  
with an exponential has been done only considering the galaxy light
beyond the 3" from the centre, because we assume  
that the light from the nucleus should not be spread out to these
distances, as the mean FWHM is 1.25" for our  
observations.

The scaling of the galactic light to match the centre of the galaxy has been done by extrapolation of the 
light profile to the centre of dEs. The scale factor C has been calculated using the following equation,

\begin{equation}
C = \frac{\sum\limits^{1}_{i= -1}F_{ i}^{g}}{ \sum\limits^{32}_{i=13}F_{i}^{g}}
\end{equation}\\
where F$^{g}$ is the flux from the best fitted galaxy profile (solid line in Fig. \ref{sef}), and i is in pixel 
scale  (i.e., 0.25") with the origin at the central peak of the observed slit profile of the galaxies. Then we 
subtract the galaxy light from the nucleus using
\begin{equation}
F^{nuc}_{\lambda} = \sum\limits^{1}_{i=-1}F^{o}_{\lambda i} - C \sum\limits^{32}_{i=13}F^{o}_{\lambda i}
\end{equation}
here, F$^{o}$ is the observed light in the frame.

Although our exponential profiles of the galaxies are in good agreement with the observed profiles (see Fig.
 \ref{fitser}), some dEs have steeper profiles than exponential \citep{Janz08} - VCC0389, VCC0929,
 VCC1167, VCC1254, VCC1348 and VCC1861 have n $\approx$2. Note, however, this finding is based on fitting a
much larger radial interval from the imaging data. In these cases, we again derived the galactic light profile 
for n = 2, which produced a better match for VCC0929. However, the calculated difference of the amount 
of galaxy light which might be left at the centre when using n = 1 was less than 30\% of the total 
central light when compared to n = 2. 
Therefore, we always used the exponential profile for scaling the galactic light to the centre for all dEs.

\begin{figure}
\centering
\includegraphics[width=8.5cm]{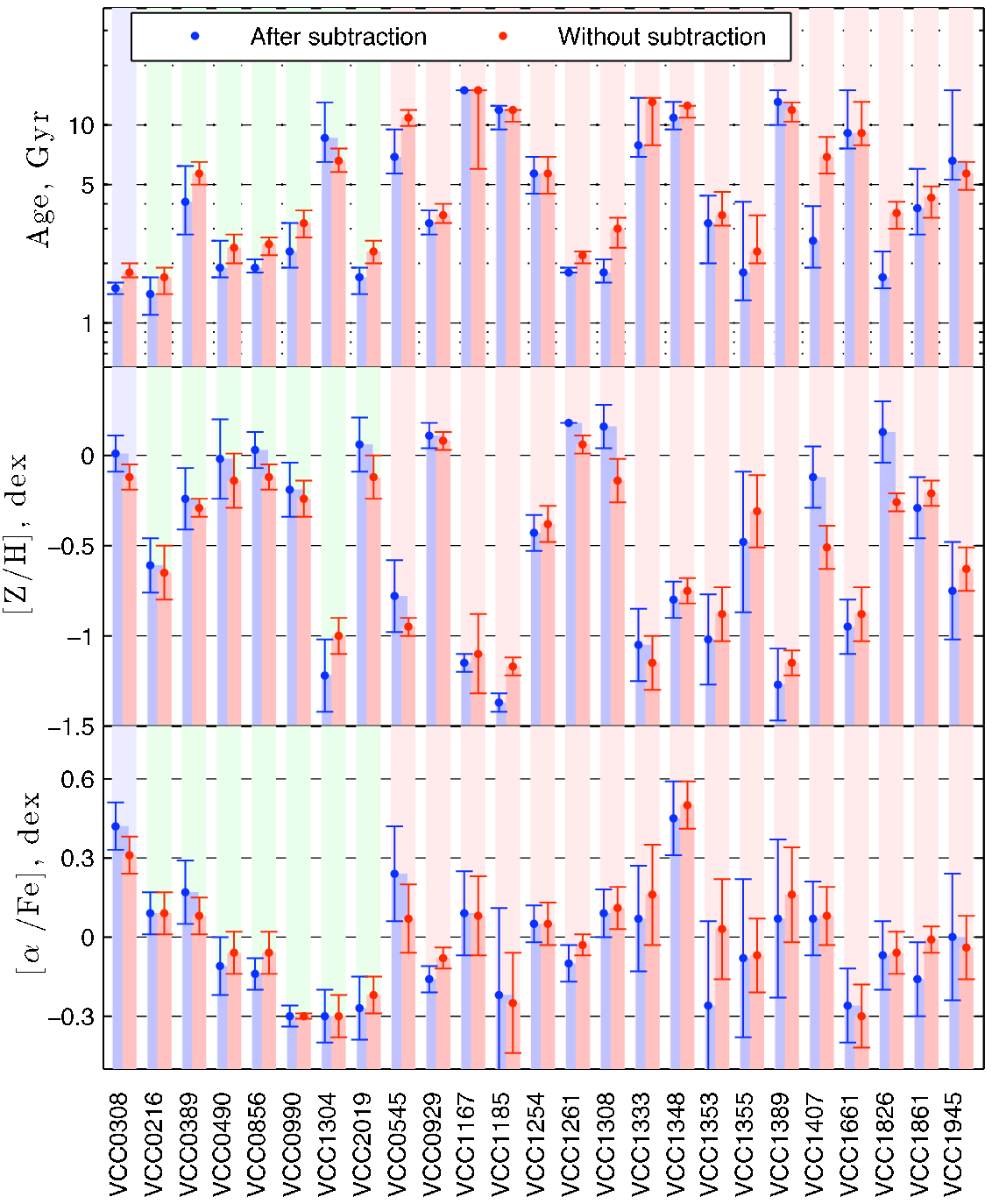}
 \caption{The comparison of the SSP-equivalent parameters after and before subtraction of galaxies'
light from the nuclei spectra.   }
 \label{diff}
\end{figure}

\begin{figure}
\centering
\includegraphics[width=8cm]{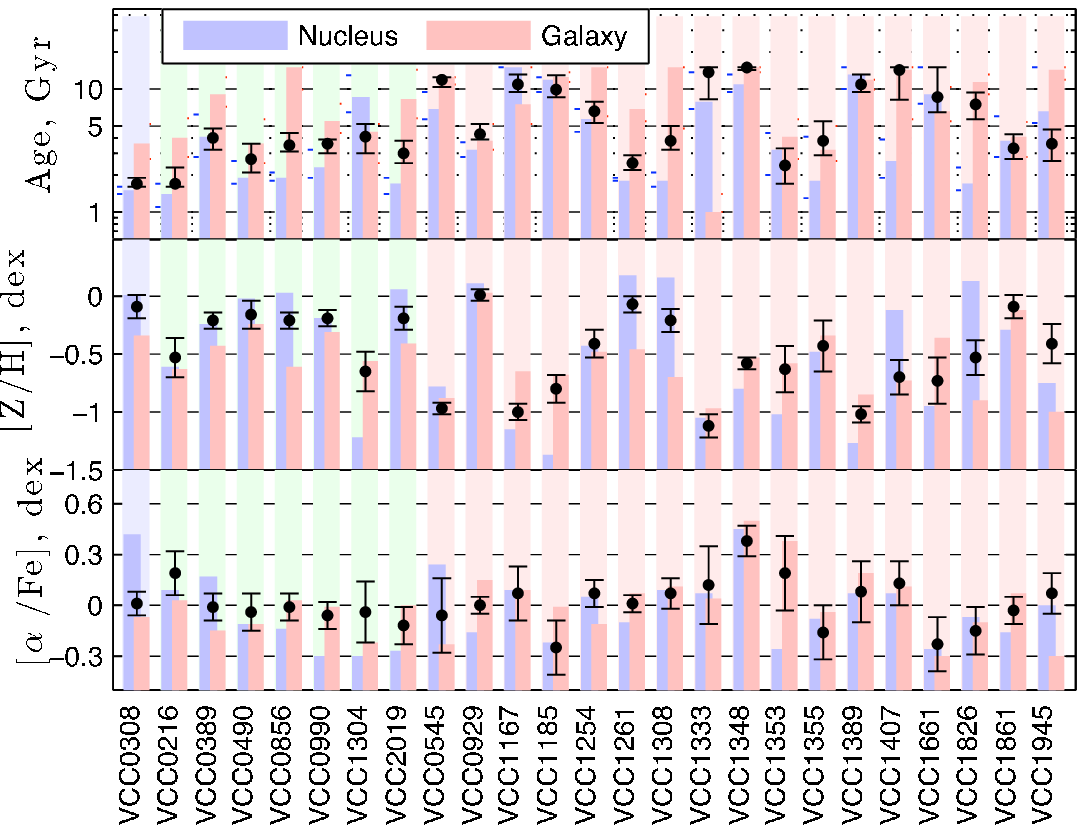}
 \caption{The comparison of the SSP-equivalent parameters of the
   galactic main bodies (red), nuclei of dEs (blue), and the result for the
   combined central light from Paper I (black), where a central
spectrum was analysed without separating nucleus and galactic main body. }
 \label{cp1st}
\end{figure}

\begin{table*}
 \centering
 \begin{minipage}{190mm}
  \caption{Measured line strength indices from the nuclei of dEs after subtraction of galactic light and corrected to the Lick system.}
  \label{indic}
\begin{tabular}{l r  |rr| rr rr rr|}
\hline
VCC	&H$\delta_{F}$&	H$\gamma_{F}$&	Fe4383			&	H$\beta$			&	Fe5015		&		Mg$_{b}$			&	Fe5270			&	Fe5335			&	Fe5406\\

no.		&	\AA			&	\AA			&	\AA	&	\AA		&	\AA &		\AA		&	\AA		&	\AA	&	\AA \\

\hline		
0216 &       4.75    $\pm$   0.27    &       3.78    $\pm$   0.24    &       0.06    $\pm$   0.64    &       3.11    $\pm$   0.28    &       3.82    $\pm$   0.62    &       1.28    $\pm$   0.31    &       1.69    $\pm$   0.35    &       0.68    $\pm$   0.39    &       1.44    $\pm$   0.28    \\
0308 &       2.95    $\pm$   0.43    &       2.48    $\pm$   0.35    &       4.08    $\pm$   0.82    &       3.31    $\pm$   0.36    &       4.49    $\pm$   0.80    &       3.06    $\pm$   0.38    &       0.84    $\pm$   0.45    &       0.82    $\pm$   0.53    &       0.72    $\pm$   0.37    \\
0389 &       2.78    $\pm$   0.50    &       -0.20   $\pm$   0.44    &       4.91    $\pm$   0.91    &       2.39    $\pm$   0.42    &       4.71    $\pm$   0.90    &       2.41    $\pm$   0.43    &       1.84    $\pm$   0.48    &       2.00    $\pm$   0.53    &       0.73    $\pm$   0.42    \\
0490 &       1.87    $\pm$   0.47    &       0.94    $\pm$   0.40    &       4.04    $\pm$   0.87    &       2.89    $\pm$   0.38    &       4.59    $\pm$   0.83    &       2.17    $\pm$   0.41    &       2.71    $\pm$   0.45    &       2.10    $\pm$   0.51    &       1.24    $\pm$   0.37    \\
0545 &       2.10    $\pm$   0.44    &       0.79    $\pm$   0.37    &       2.77    $\pm$   0.86    &       2.84    $\pm$   0.37    &       2.16    $\pm$   0.85    &       2.00    $\pm$   0.40    &       1.36    $\pm$   0.45    &       1.00    $\pm$   0.51    &       1.21    $\pm$   0.38    \\
0725$^{a}$   &       2.01    $\pm$   0.57    &       2.28    $\pm$   0.50    &       2.16    $\pm$   1.22    &       3.55    $\pm$   0.52    &       2.55    $\pm$   1.24    &       1.44    $\pm$   0.57    &       1.16    $\pm$   0.67    &       0.54    $\pm$   0.77    &       0.39    $\pm$   0.56    \\
0856 &       1.76    $\pm$   0.27    &       1.27    $\pm$   0.22    &       4.81    $\pm$   0.49    &       2.02    $\pm$   0.23    &       5.17    $\pm$   0.48    &       2.35    $\pm$   0.24    &       2.69    $\pm$   0.26    &       2.29    $\pm$   0.29    &       1.40    $\pm$   0.22    \\
0929 &       0.04    $\pm$   0.31    &       -0.28   $\pm$   0.25    &       5.45    $\pm$   0.51    &       2.35    $\pm$   0.24    &       5.68    $\pm$   0.50    &       2.82    $\pm$   0.25    &       2.41    $\pm$   0.28    &       3.15    $\pm$   0.29    &       1.83    $\pm$   0.22    \\
0990 &       1.02    $\pm$   0.59    &       1.01    $\pm$   0.45    &       3.03    $\pm$   0.97    &       2.94    $\pm$   0.39    &       5.80    $\pm$   0.82    &       1.33    $\pm$   0.41    &       2.71    $\pm$   0.44    &       1.25    $\pm$   0.51    &       2.35    $\pm$   0.36    \\
1167 &       3.05    $\pm$   0.30    &       1.45    $\pm$   0.28    &       1.55    $\pm$   0.62    &       2.58    $\pm$   0.27    &       3.56    $\pm$   0.61    &       1.38    $\pm$   0.30    &       1.50    $\pm$   0.33    &       1.13    $\pm$   0.38    &       0.71    $\pm$   0.28    \\
1185 &       1.37    $\pm$   0.44    &       2.05    $\pm$   0.34    &       2.03    $\pm$   0.85    &       2.17    $\pm$   0.41    &       2.36    $\pm$   0.90    &       1.04    $\pm$   0.43    &       1.08    $\pm$   0.48    &       0.48    $\pm$   0.54    &       0.44    $\pm$   0.41    \\
1254 &       1.60    $\pm$   0.23    &       0.60    $\pm$   0.20    &       3.36    $\pm$   0.43    &       1.93    $\pm$   0.19    &       4.14    $\pm$   0.42    &       2.31    $\pm$   0.20    &       1.85    $\pm$   0.23    &       2.06    $\pm$   0.25    &       1.28    $\pm$   0.19    \\
1261 &       1.58    $\pm$   0.32    &       0.45    $\pm$   0.27    &       4.07    $\pm$   0.59    &       2.77    $\pm$   0.26    &       6.05    $\pm$   0.55    &       2.76    $\pm$   0.27    &       3.27    $\pm$   0.30    &       2.95    $\pm$   0.34    &       1.42    $\pm$   0.25    \\
1304 &       1.94    $\pm$   0.41    &       1.75    $\pm$   0.35    &       1.41    $\pm$   0.85    &       2.46    $\pm$   0.37    &       2.48    $\pm$   0.85    &       0.82    $\pm$   0.42    &       1.70    $\pm$   0.45    &       1.57    $\pm$   0.50    &       0.70    $\pm$   0.39    \\
1308 &       1.39    $\pm$   0.58    &       1.28    $\pm$   0.46    &       5.30    $\pm$   1.00    &       2.55    $\pm$   0.45    &       1.11    $\pm$   1.10    &       3.11    $\pm$   0.47    &       1.44    $\pm$   0.55    &       2.24    $\pm$   0.61    &       1.52    $\pm$   0.46    \\
1333 &       1.99    $\pm$   0.34    &       1.72    $\pm$   0.29    &       3.00    $\pm$   0.70    &       2.20    $\pm$   0.31    &       3.48    $\pm$   0.69    &       1.61    $\pm$   0.34    &       0.81    $\pm$   0.38    &       0.99    $\pm$   0.43    &       0.54    $\pm$   0.32    \\
1348 &       1.68    $\pm$   0.36    &       0.29    $\pm$   0.32    &       2.40    $\pm$   0.69    &       2.16    $\pm$   0.31    &       3.07    $\pm$   0.68    &       2.26    $\pm$   0.32    &       1.74    $\pm$   0.36    &       0.85    $\pm$   0.42    &       0.50    $\pm$   0.31    \\
1353 &       3.84    $\pm$   0.47    &       3.50    $\pm$   0.44    &       1.32    $\pm$   1.10    &       3.00    $\pm$   0.48    &       3.71    $\pm$   1.06    &       1.13    $\pm$   0.52    &       0.59    $\pm$   0.60    &       1.62    $\pm$   0.65    &       0.87    $\pm$   0.51    \\
1355 &       4.12    $\pm$   0.75    &       1.72    $\pm$   0.76    &       2.67    $\pm$   1.76    &       2.75    $\pm$   0.76    &       2.89    $\pm$   1.74    &       1.44    $\pm$   0.80    &       2.42    $\pm$   0.89    &       1.66    $\pm$   1.02    &       1.42    $\pm$   0.75    \\
1389 &       3.15    $\pm$   0.47    &       1.27    $\pm$   0.44    &       1.47    $\pm$   1.01    &       1.65    $\pm$   0.44    &       2.75    $\pm$   0.98    &       1.37    $\pm$   0.46    &       0.72    $\pm$   0.52    &       1.33    $\pm$   0.58    &       0.97    $\pm$   0.42    \\
1407 &       1.61    $\pm$   0.70    &       0.69    $\pm$   0.54    &       2.96    $\pm$   1.25    &       2.52    $\pm$   0.50    &       4.47    $\pm$   1.05    &       2.64    $\pm$   0.50    &       1.43    $\pm$   0.57    &       2.42    $\pm$   0.61    &       1.92    $\pm$   0.46    \\
1661 &       2.11    $\pm$   0.41    &       0.35    $\pm$   0.36    &       3.02    $\pm$   0.82    &       2.49    $\pm$   0.36    &       3.33    $\pm$   0.82    &       1.09    $\pm$   0.40    &       1.60    $\pm$   0.44    &       1.12    $\pm$   0.50    &       1.33    $\pm$   0.36    \\
1826 &       1.78    $\pm$   1.62    &       0.75    $\pm$   0.92    &       5.46    $\pm$   1.59    &       3.01    $\pm$   0.56    &       2.78    $\pm$   1.26    &       2.65    $\pm$   0.55    &       3.09    $\pm$   0.59    &       2.34    $\pm$   0.67    &       1.56    $\pm$   0.49    \\
1861 &       1.84    $\pm$   0.50    &       0.52    $\pm$   0.44    &       4.01    $\pm$   0.93    &       1.78    $\pm$   0.43    &       4.63    $\pm$   0.90    &       2.00    $\pm$   0.44    &       2.62    $\pm$   0.48    &       1.92    $\pm$   0.55    &       1.71    $\pm$   0.42    \\
1945 &       2.04    $\pm$   0.43    &       0.88    $\pm$   0.37    &       2.41    $\pm$   0.87    &       2.29    $\pm$   0.40    &       4.32    $\pm$   0.89    &       1.66    $\pm$   0.43    &       1.09    $\pm$   0.50    &       1.17    $\pm$   0.57    &       1.58    $\pm$   0.41    \\
2019 &       1.84    $\pm$   0.52    &       1.24    $\pm$   0.40    &       4.44    $\pm$   0.91    &       2.95    $\pm$   0.41    &       4.45    $\pm$   0.90    &       2.00    $\pm$   0.45    &       2.99    $\pm$   0.49    &       2.34    $\pm$   0.56    &       1.71    $\pm$   0.43    \\
\hline
 \end{tabular}
$^{a}$without subtraction of galactic light. 
\end{minipage}

\end{table*}

\begin{table*}
 \centering
  \caption{Measured line strength indices from the galactic main body
    of dEs (i.e., 3 to 8 arcsec radial interval) and corrected to the Lick system.}
  \label{indic}
\begin{tabular}{l r  |rr| rr rr rr|}

\hline
VCC		&	H$\delta_{F}$			&	H$\gamma_{F}$			&	Fe4383			&	H$\beta$			&	Fe5015		&		Mg$_{b}$			&	Fe5270			&	Fe5335			&	Fe5406\\

no.		&	\AA			&	\AA			&	\AA	&	\AA		&	\AA &		\AA		&	\AA		&	\AA	&	\AA \\

\hline
0216 &       2.46    $\pm$   0.38    &       1.87    $\pm$   0.34    &       3.22    $\pm$   0.80    &       2.35    $\pm$   0.36    &       3.13    $\pm$   0.84    &       2.02    $\pm$   0.40    &       1.61    $\pm$   0.46    &       1.35    $\pm$   0.51    &       1.16    $\pm$   0.38    \\
0308 &       1.56    $\pm$   0.38    &       0.85    $\pm$   0.34    &       3.30    $\pm$   0.78    &       2.46    $\pm$   0.36    &       4.43    $\pm$   0.81    &       2.00    $\pm$   0.39    &       2.37    $\pm$   0.44    &       2.06    $\pm$   0.50    &       1.07    $\pm$   0.37    \\
0389 &       -0.57   $\pm$   0.43    &       -0.06   $\pm$   0.37    &       3.26    $\pm$   0.79    &       2.49    $\pm$   0.36    &       4.10    $\pm$   0.81    &       2.22    $\pm$   0.38    &       2.48    $\pm$   0.43    &       1.89    $\pm$   0.48    &       1.21    $\pm$   0.37    \\
0490 &       1.26    $\pm$   0.53    &       0.64    $\pm$   0.48    &       5.05    $\pm$   1.04    &       2.26    $\pm$   0.49    &       4.38    $\pm$   1.10    &       2.23    $\pm$   0.53    &       2.35    $\pm$   0.60    &       1.87    $\pm$   0.67    &       1.19    $\pm$   0.49    \\
0545 &       -0.05   $\pm$   0.43    &       0.09    $\pm$   0.37    &       2.51    $\pm$   0.83    &       2.13    $\pm$   0.38    &       3.82    $\pm$   0.85    &       1.26    $\pm$   0.42    &       0.78    $\pm$   0.48    &       0.86    $\pm$   0.53    &       1.16    $\pm$   0.39    \\
0856 &       1.39    $\pm$   0.35    &       -0.48   $\pm$   0.32    &       3.80    $\pm$   0.70    &       1.94    $\pm$   0.33    &       3.84    $\pm$   0.73    &       2.11    $\pm$   0.35    &       2.07    $\pm$   0.39    &       1.84    $\pm$   0.44    &       1.23    $\pm$   0.33    \\
0929 &       1.71    $\pm$   0.32    &       -0.66   $\pm$   0.31    &       5.48    $\pm$   0.62    &       2.25    $\pm$   0.30    &       4.40    $\pm$   0.66    &       3.21    $\pm$   0.31    &       2.78    $\pm$   0.36    &       1.75    $\pm$   0.40    &       1.43    $\pm$   0.30    \\
0990 &       0.88    $\pm$   0.35    &       0.36    $\pm$   0.31    &       4.18    $\pm$   0.66    &       2.21    $\pm$   0.30    &       4.52    $\pm$   0.70    &       2.27    $\pm$   0.32    &       2.12    $\pm$   0.37    &       1.68    $\pm$   0.42    &       1.31    $\pm$   0.32    \\
1167 &       1.79    $\pm$   0.41    &       0.37    $\pm$   0.39    &       3.41    $\pm$   0.84    &       2.59    $\pm$   0.39    &       3.07    $\pm$   0.91    &       2.08    $\pm$   0.43    &       1.75    $\pm$   0.48    &       1.01    $\pm$   0.58    &       1.41    $\pm$   0.40    \\
1185 &       1.50    $\pm$   0.32    &       -0.88   $\pm$   0.31    &       3.54    $\pm$   0.65    &       1.53    $\pm$   0.32    &       3.48    $\pm$   0.69    &       1.59    $\pm$   0.35    &       1.08    $\pm$   0.39    &       2.12    $\pm$   0.42    &       1.19    $\pm$   0.32    \\
1254 &       0.60    $\pm$   0.45    &       -0.18   $\pm$   0.39    &       5.24    $\pm$   0.83    &       1.20    $\pm$   0.41    &       3.84    $\pm$   0.91    &       2.86    $\pm$   0.42    &       1.86    $\pm$   0.50    &       2.48    $\pm$   0.55    &       2.09    $\pm$   0.40    \\
1261 &       1.20    $\pm$   0.27    &       0.40    $\pm$   0.25    &       3.61    $\pm$   0.56    &       2.03    $\pm$   0.26    &       3.66    $\pm$   0.57    &       2.38    $\pm$   0.28    &       2.16    $\pm$   0.31    &       1.51    $\pm$   0.36    &       1.37    $\pm$   0.26    \\
1304 &       2.31    $\pm$   0.37    &       0.97    $\pm$   0.36    &       3.88    $\pm$   0.78    &       2.35    $\pm$   0.37    &       4.14    $\pm$   0.82    &       1.65    $\pm$   0.40    &       2.31    $\pm$   0.44    &       1.90    $\pm$   0.49    &       1.06    $\pm$   0.38    \\
1308 &       1.72    $\pm$   0.43    &       0.15    $\pm$   0.40    &       2.38    $\pm$   0.91    &       1.94    $\pm$   0.41    &       4.91    $\pm$   0.93    &       2.24    $\pm$   0.43    &       1.86    $\pm$   0.50    &       2.15    $\pm$   0.56    &       1.16    $\pm$   0.43    \\
1333 &       6.16    $\pm$   0.39    &       1.44    $\pm$   0.50    &       1.13    $\pm$   1.04    &       3.69    $\pm$   0.44    &       1.94    $\pm$   1.08    &       1.60    $\pm$   0.51    &       1.48    $\pm$   0.57    &       0.97    $\pm$   0.67    &       1.29    $\pm$   0.47    \\
1348 &       2.56    $\pm$   0.46    &       -1.30   $\pm$   0.49    &       3.50    $\pm$   0.98    &       1.94    $\pm$   0.49    &       3.15    $\pm$   1.06    &       3.14    $\pm$   0.48    &       0.39    $\pm$   0.59    &       1.00    $\pm$   0.66    &       1.50    $\pm$   0.47    \\
1353 &       2.37    $\pm$   0.37    &       1.52    $\pm$   0.35    &       0.42    $\pm$   0.85    &       2.78    $\pm$   0.38    &       3.08    $\pm$   0.86    &       2.21    $\pm$   0.40    &       1.78    $\pm$   0.46    &       1.14    $\pm$   0.52    &       1.43    $\pm$   0.40    \\
1355 &       1.86    $\pm$   0.43    &       1.08    $\pm$   0.40    &       3.05    $\pm$   0.90    &       2.45    $\pm$   0.43    &       4.65    $\pm$   0.97    &       1.93    $\pm$   0.46    &       2.03    $\pm$   0.52    &       1.46    $\pm$   0.59    &       1.63    $\pm$   0.44    \\
1389 &       1.12    $\pm$   0.39    &       0.18    $\pm$   0.37    &       2.32    $\pm$   0.80    &       2.15    $\pm$   0.38    &       2.77    $\pm$   0.90    &       1.83    $\pm$   0.42    &       1.75    $\pm$   0.47    &       1.19    $\pm$   0.54    &       0.29    $\pm$   0.41    \\
1407 &       1.40    $\pm$   0.39    &       0.55    $\pm$   0.34    &       3.33    $\pm$   0.78    &       1.90    $\pm$   0.36    &       3.87    $\pm$   0.80    &       2.34    $\pm$   0.38    &       2.03    $\pm$   0.43    &       1.51    $\pm$   0.49    &       0.91    $\pm$   0.37    \\
1661 &       -0.04   $\pm$   0.45    &       0.02    $\pm$   0.40    &       4.31    $\pm$   0.89    &       2.17    $\pm$   0.42    &       5.75    $\pm$   0.96    &       1.30    $\pm$   0.46    &       1.10    $\pm$   0.53    &       2.63    $\pm$   0.57    &       1.63    $\pm$   0.42    \\
1826 &       1.85    $\pm$   0.35    &       -0.13   $\pm$   0.34    &       1.10    $\pm$   0.75    &       2.02    $\pm$   0.36    &       4.58    $\pm$   0.82    &       1.34    $\pm$   0.40    &       1.97    $\pm$   0.45    &       1.91    $\pm$   0.50    &       1.15    $\pm$   0.37    \\
1861 &       1.32    $\pm$   0.39    &       -0.02   $\pm$   0.36    &       4.58    $\pm$   0.74    &       2.11    $\pm$   0.35    &       4.68    $\pm$   0.77    &       2.88    $\pm$   0.37    &       1.92    $\pm$   0.43    &       2.32    $\pm$   0.47    &       1.56    $\pm$   0.37    \\
1945 &       1.96    $\pm$   0.35    &       0.94    $\pm$   0.33    &       6.20    $\pm$   0.70    &       1.66    $\pm$   0.36    &       2.89    $\pm$   0.81    &       1.59    $\pm$   0.39    &       1.46    $\pm$   0.44    &       2.09    $\pm$   0.49    &       1.23    $\pm$   0.36    \\
2019 &       0.68    $\pm$   0.43    &       0.02    $\pm$   0.37    &       3.97    $\pm$   0.82    &       1.94    $\pm$   0.39    &       4.63    $\pm$   0.89    &       2.35    $\pm$   0.42    &       2.12    $\pm$   0.49    &       1.55    $\pm$   0.56    &       1.36    $\pm$   0.42    \\
\hline
 \end{tabular}
\end{table*}


\section{Extraction of SSP parameters}\label{degen}
\begin{figure}
\includegraphics[width=7cm]{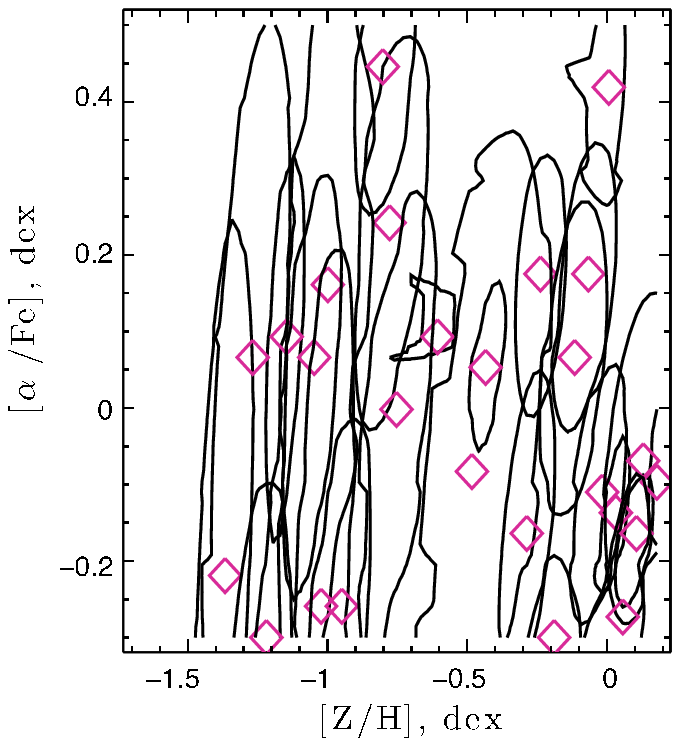}
\includegraphics[width=7cm]{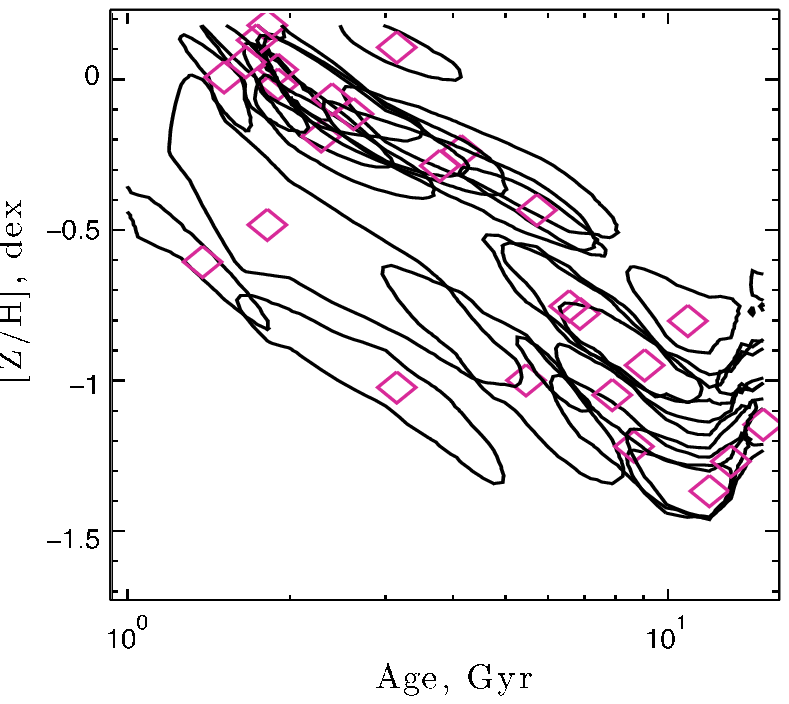}
\includegraphics[width=7cm]{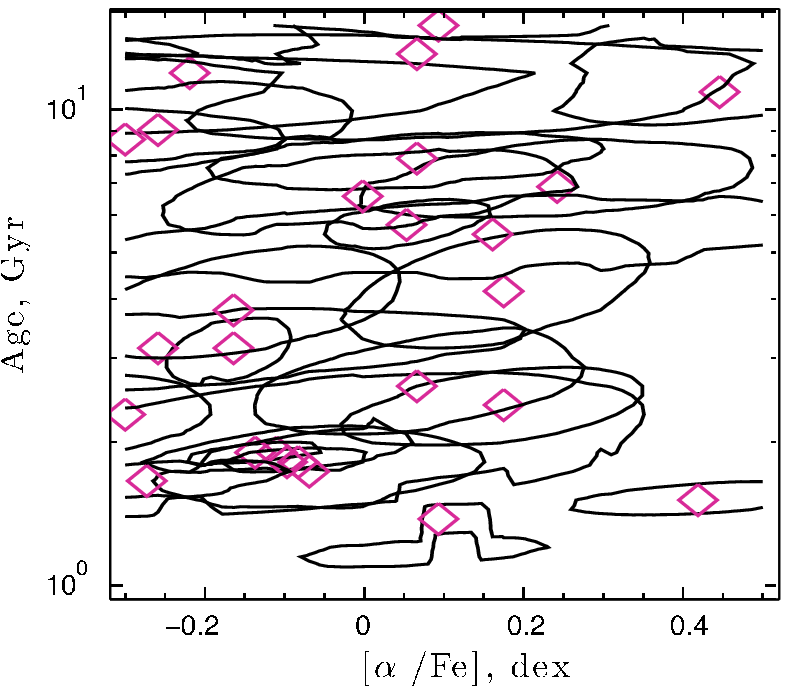}
 \caption{Examples of $\Delta\chi^{2}$ contours in different projection planes of age, metallicity and [$\alpha$/Fe]-abundance space. }
 \label{chi}
 \end{figure}

It is well known that the age-metallicity degeneracy is a difficult problem to estimate galaxy 
age and metallicity. However, there are several different methods have been suggested to cope with this complication. By using the large number of indices and adopting the technique of \cite{Proctor02}, the effect of this degeneracy on the estimates of SSP parameters can be minimized. Fig. \ref{chi}, shows examples of the of $\Delta\chi^{2}$ contours obtained with the method we have used to derive the SSP parameters, indicating the minimum with a diamond symbol. The contours are drawn with $\Delta\chi^{2}$ = 2.3 (i.e., errors including 2 degrees of freedom \citep[][Section 15.6]{Press92}). This shows that the typical 1$\sigma$ uncertainties we obtain on the SSP paramters are of the order of 0.1 dex. The effect of the age-metallicity degeneracy   \citep[e.g.][]{Worthey94a} can be recognized in the tilt of the contours in the age Vs metallicity plot.

\end{document}